\documentclass[prl,reprint,twocolumn,showpacs,superscriptaddress,floatfix,aps,10pt]{revtex4-2}
\usepackage{amsmath,amsthm,amssymb}
\usepackage{dcolumn,bm,hyperref,braket}
\usepackage{graphicx}
\usepackage{subfigure,verbatim}
\usepackage{xcolor}
\usepackage{soul}
\usepackage{ulem}
\definecolor{LinkColor}{RGB}{46,48,146}
\usepackage{hyperref}
\hypersetup{
colorlinks=true,
citecolor=LinkColor,
linkcolor=LinkColor,
urlcolor=LinkColor
}

\begin{document}
\title{Magnetism and Topology from Circularly Polarized Phonon Floquet Engineering}

\author{Dapeng Yao}
%\email{dapeng.yao@riken.jp}
\affiliation{RIKEN Center for Emergent Matter Science (CEMS), 2-1 Hirosawa, Wako, Saitama 351-0198, Japan}

\author{Tiantian Zhang}
\affiliation{Institute of Theoretical Physics, Chinese Academy of Sciences, Beijing 100190, China}

\author{Takashi Oka}
\affiliation{The Institute for Solid State Physics, The University of Tokyo, Kashiwa, Chiba 277-8581, Japan}

\author{Takehito Yokoyama}
\affiliation{Department of Physics, Institute of Science Tokyo, 2-12-1 Ookayama, Meguro-ku, Tokyo 152-8551, Japan}

\begin{abstract}
We theoretically show that circularly polarized phonons induce electronic magnetization and drive a topological phase transition via phonon Floquet engineering. Considering the electronic states modulated by circularly polarized phonons on a honeycomb lattice, we show that such lattice dynamics generates an effective next-nearest-neighbor electron hopping, leading to a Haldane-type mass term.
Circularly polarized phonon breaks time-reversal symmetry (TRS) and opens a gap at valley points, undergoing phase transition from a trivial insulator to a Chern insulator. 
Moreover, the orbital and spin magnetizations emerge due to the breaking of TRS. Our results show that circularly polarized phonons serve as an effective magnetic field to engineer magnetism and topology, offering new opportunities for phonon Floquet approaches.
\end{abstract}
\maketitle

Recent advances in chiral or axial phonons with circularly polarized modes~\cite{Zhang2014,Zhang2015,Zhu2018,Juraschek2019,Zhang2022,Zhang2023,Ishito2023,Ueda2023,Ohe2024,Zhang2025,Zhang2025NC,Juraschek2025,Natalia2025,Che2025,Yang2025,S_Zhang2026} has opened a new avenue for angular momentum-transfer effect between phonons and electrons~\cite{Juraschek2020,Juraschek2022,Luo2023,Kim2023,Geilhufe2023,YaoAPL2024,Funato2024,Yokoyama2024,Shabala2024,Yokoyama2025,Nishimura2025,Yokoyama2026}. Since such phonons dynamically modulate electronic degrees of freedom, periodically driven electron systems can be treated within adiabatic approximation~\cite{Berry1984} which requires that the phonon energy scale is much smaller than the electron bandwidth. Within adiabatic evolutions, electrons acquire a geometric phase~\cite{Trifunovic2019}, leading to a dynamical electric current~\cite{Yao2022}, spin magnetization~\cite{Hamada2020,Ren2024,Yao2024,Yao2025,Royo2025}, and orbital magnetization~\cite{Ren2021,Xiao2021,YaoOAM2025,Sato2025,Pezp2026}.
However, adiabatic treatment breaks down when electron gap closes due to phonon dynamics.
In the opposite limit, Floquet theory provides a powerful framework to engineer quantum materials in the ultrafast regime~\cite{Floquet1883,Shirley1965}. It has been applied to a wide range of nonequilibrium phenomena, including light-induced band topology~\cite{Oka2009,Kitagawa2010,Kitagawa2011,Lindner2011,Dag2022,Oka2019,Rudner2020,Harper2020,Wang2026}, ultrafast spintronics~\cite{Sentef2015,Kibis2022,Neufeld2023,Tanaka2024}, phonon Floquet engineering~\cite{Murakami2017,Shin2018,Hubener2018,Chaudhary2020,Klebl2025}.
By contrast, the Floquet engineering of electronic topology and magnetism mediated by circularly polarized phonons remains less explored.

\begin{figure}
\begin{center}
\includegraphics[width=8cm]{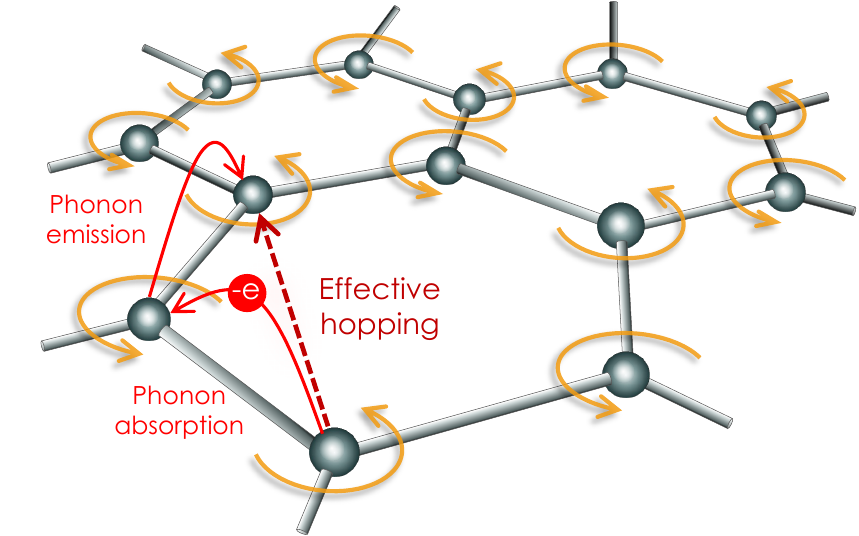}
\caption{Schematic of the phonon-driven electron systems on a two-dimensional honeycomb lattice, where a circularly polarized phonon in optical branches at the Brillouin-zone center is excited. Electron-phonon coupling enables phonon-assisted nearest-neighbor hopping via phonon emission and absorption processes, which generates an effective next-nearest-neighbor hopping. This process drives the system to a topological phase with a nonzero Chern number. Spin and orbital magnetizations emerge due to the breaking of time-reversal symmetry by circularly polarized phonons.}
\label{fig1}
\end{center}
\end{figure}

In this Letter, we show that magnetism and topology emerge via circularly polarized phonon Floquet engineering. We consider a microscopic mechanism of electron-phonon coupling on a honeycomb lattice, i.e., electrons are modulated by circularly polarized phonons and phonon dynamics is imprinted in electronic states. Here we focus on the ultrafast regime, where electrons are driven by the circularly polarized phonon at high frequency. Unlike the Floquet approaches applied in momentum space~\cite{Shin2018,Hubener2018,Chaudhary2020,Klebl2025}, we formulate the Floquet engineering in real spaces, where such phonons generate an effective next-nearest-neighbor (NNN) electron hopping via the phonon emission and absorption as shown in Fig.~\ref{fig1}.
This process leads to a Haldane-type mass term~\cite{Haldane1985}, driving a transition from a trivial insulator to a Chern insulator due to the phonon-induced gap closing. In the presence of spin-orbital coupling (SOC), further gap closings occur, leading to a change of the Chern number in topological phases.
Furthermore, the orbital and spin magnetizations appear due to the breaking of time-reversal symmetry (TRS). Our results show that circularly polarized phonons serve as an effective magnetic field to dynamically engineer magnetism and topology, offering a new route toward phonon Floquet engineering.

\textit{Phonon-driven electron system.}---
We assume that the circularly polarized phonons are driven by light~\cite{Nova2017} and neglect the coupling between electrons and light. Since the relaxation timescale of lattice dynamics is typically much slower than the response of electrons to light, the phonon-driven Floquet effect can be distinguished from that driven by light. Therefore, here we focus on phonon-driven electron systems.
We begin with the tight-binding model of the two-dimensional honeycomb lattice, where circularly polarized phonons modulate electronic states. In the absence of phonons, the real-space electron Hamiltonian reads
\begin{equation}
H_0=t_0\sum_{\braket{ij}}c_{i}^{\dagger}c_j+\sum_{i}\delta_ic^{\dagger}_{i}c_{i}+i\lambda_{\text{R}}\sum_{\braket{ij}}c^{\dagger}_i(\bm s\times\hat{\bm d}_{ij})_zc_j,
\end{equation}
where $c_i=(c_{i\uparrow},c_{i\downarrow})^{\mathsf{T}}$ $[c^{\dagger}_i=(c^{\dagger}_{i\uparrow},c^{\dagger}_{i\downarrow})]$ is the annihilation (creation) operator of electron at the $i$th site. The first term represents the nearest-neighbor (NN) electron hopping with the amplitude $t_0$, and the second term describes a staggered on-site potential $\delta_{i}=\pm\delta$ for the A(B) sublattice. The third term denotes the Rashba SOC with the parameter $\lambda_{\text{R}}$, where $\hat{\bm d}_{ij}=\frac{\bm d_{ij}}{|\bm d_{ij}|}$ is the unit vector from the site $i$ to $j$, and $\bm s$ is the Pauli-matrices vector standing for electron spins.

At the phonon Brillouin-zone center, the superposition of the doubly degenerate optical phonons form circularly polarized modes, classified as counterclockwise and clockwise modes~\cite{Zhang2015}. Let $\bm u_i(t)$ and $\bm u_j(t)$ be the time-dependent displacements of the atoms at the sites $i$ and $j$, respectively. When the circularly polarized phonons are present, the bond vector changes into $\bm d_{ij}\rightarrow\bm d_{ij}+\bm u_{ij}(t)$ with $\bm u_{ij}(t)=\bm u_{j}(t)-\bm u_{i}(t)$ being the relative displacement of the adjacent sites, and the NN atomic distance accordingly changes to $a_0\rightarrow a_0+\frac{1}{a_0}\bm u_{ij}(t)\cdot\bm d_{ij}$. For simplicity, we assume that the electron hoppings only involve the isotropic atomic orbital, and the hopping amplitude is modulated as $t_0\rightarrow t_0-\frac{t_0}{a_0^2}\bm u_{ij}(t)\cdot\bm d_{ij}$.
Meanwhile, $\frac{1}{|\bm d_{ij}|}$ in the Rashba SOC term changes to
$\frac{1}{|\bm d_{ij}|}\rightarrow\frac{1}{|\bm d_{ij}+\bm u_{ij}(t)|}=\frac{1}{a_0}-\frac{1}{a_0^3}\bm u_{ij}(t)\cdot\bm d_{ij}+\mathcal{O}[\bm u_{ij}^2(t)]$. As a result, the electron-phonon coupling is given by
\begin{align}
H_{\text{ep}}(t)=\sum_{\braket{ij}}c^{\dagger}_i\Big[&-\frac{t_0}{a_0^2}\bm u_{ij}(t)\cdot\bm d_{ij}+i\frac{\lambda_{\text{R}}}{a_0}[\bm s\times\bm u_{ij}(t)]_z\nonumber\\
&-i\frac{\lambda_{\text{R}}}{a_0^3}\bm u_{ij}(t)\cdot\bm d_{ij}(\bm s\times\bm d_{ij})_z\Big]c_j,
\end{align}
which is considered with respect to the first order of the phonon relative displacement.

\textit{Floquet picture.}---
We take the counterclockwise mode as an example, where the relative displacement in the optical branch at the zone center is represented by $\bm u_{ij}(t)=u_{ij}(\cos\Omega t,\sin\Omega t)$ with $u_{ij}=u_j-u_i$. For the phonon modes at $\Gamma$ point, the same species of atoms rotate in phase, yielding the relative rotational amplitude $u_{ij}=\pm u_r$ with $u_r\equiv u_{\text{B}}-u_{\text{A}}$.
The time-dependent Hamiltonian is then decomposed in terms of the two Fourier modes: $H_{\text{ep}}(t)=H_1e^{i\Omega t}+H_{-1}e^{-i\Omega t}$. We can express the Fourier component as $H_{1}=\sum_{\braket{ij}}c^{\dagger}_i\sum_{\alpha=0,x,y}J_{ij,\alpha}^{(1)}s_{\alpha}c_j$, where $J^{(1)}_{ij,0}=-\frac{t_0u_{ij}}{2a_0}e^{-i\theta_{ij}}$, $J^{(1)}_{ij,x}=\frac{\lambda_{\text{R}}u_{ij}}{4a_0}(1+e^{-i2\theta_{ij}})$, and $J^{(1)}_{ij,y}=-i\frac{\lambda_{\text{R}}u_{ij}}{4a_0}(1-e^{-i2\theta_{ij}})$ with $\theta_{ij}=\arctan(\frac{d_{ij}^y}{d_{ij}^x})$ being the azimuthal angle of $\bm d_{ij}$. Here the Fourier components satisfy $H_{-1}=H^{\dagger}_{1}$, leading to $J^{(-1)}_{ij}=(J^{(1)}_{ji})^*$ due to the Hermiticity of Hamiltonian.

In our phonon-driven electron systems, adiabatic treatment breaks down when the driving frequency $\Omega$ is much larger than the scale of the electron energy gap $\delta/\hbar$. Instead, in the high-frequency regime, an effective Floquet Hamiltonian can be obtained by using the van-Vleck type expansion with respect to $\Omega^{-1}$ as~\cite{Kitagawa2011,Goldman2014,Bukov2015,Eckardt2017}
\begin{align}
H_{\text{eff}}=H_0+\frac{1}{2\hbar\Omega}[H_{-1},H_1],
\end{align}
where the second term describing the phonon dynamics under the high-frequency expansion reads
\begin{align}\label{H_Omega}
\frac{1}{2\hbar\Omega}\sum_{\langle\!\langle ij\rangle\!\rangle}c_i^{\dagger}\Big[\tilde Ai\nu_{ij}s_0-\tilde B\cos{\frac{\phi_{ij}}{2}}s_x+\tilde B\sin\frac{\phi_{ij}}{2}s_y+\tilde Cs_z\Big]c_j,
\end{align}
with the coefficients $\tilde A\equiv\frac{\sqrt{3}}{8}(2t_0^2-\lambda_{\text{R}}^2)\left(\frac{u_r}{a_0}\right)^2$, $\tilde B\equiv\frac{t_0\lambda_{\text{R}}}{4}\left(\frac{u_r}{a_0}\right)^2$, and $\tilde C\equiv\frac{\lambda_{\text{R}}^2}{8}\left(\frac{u_r}{a_0}\right)^2$~\cite{SM}.
We notice that the first term in Eq.~(\ref{H_Omega}) shows an effective complex NNN hopping from the two-step phonon-mediated process as illustrated in Fig.~\ref{fig1}. Here, $\phi_{ij}\equiv\theta_{ki}+\theta_{kj}$ characterizes the two-step path connecting the NNN sites $i$ and $j$ via the intermediate site $k$, and $\nu_{ij}=1(-1)$ denotes the hopping with the clockwise (counterclockwise) path from the site $i$ to $j$ on the honeycomb lattice.

Here we choose the Dirac matrices as $\Gamma^{1,2,3,4,5}=(\sigma_xs_0,\sigma_zs_0,\sigma_ys_x,\sigma_ys_y,\sigma_ys_z)$, where the Pauli matrices $\sigma_a$ and $s_a$ denote the Pauli matrices in sublattice index and electron spin, respectively. Their ten commutators are given by $\Gamma^{ab}=\frac{1}{2i}[\Gamma^a,\Gamma^b]$. The Bloch Hamiltonian $\mathcal{H}_{\text{eff}}(\bm k)$ is then written as a linear combination of these Dirac matrices:
\begin{align}
\mathcal{H}_{\text{eff}}(\bm k)=\sum_{a=1}^4R_a(\bm k)\Gamma^a+\sum_{a<b=1}^{4}R_{ab}(\bm k)\Gamma^{ab},
\end{align}
where $R_1=t_0(1+2\cos x\cos y)$, $R_{12}=-2t_0\cos x\sin y$, $R_4=-\sqrt{3}\lambda_{\text{R}}\sin x\sin y$, $R_{23}=-\lambda_{\text{R}}\cos x\sin y$, $R_{24}=\sqrt{3}\lambda_{\text{R}}\sin x\cos y$, $R_3=\lambda_{\text{R}}(1-\cos x\cos y)$, $R_2=\delta+\frac{\tilde{A}}{\hbar\Omega}\sin x(\cos x-\cos y)$, $R_{13}=\frac{\sqrt{3}\tilde{B}}{\hbar\Omega}\cos x\cos y$, $R_{14}=\frac{\tilde{B}}{\hbar\Omega}(\sin x\sin y+\cos 2x)$, and $R_{34}=\frac{\tilde{C}}{\hbar\Omega}(2\cos x\cos y+\cos 2x)$ with $x=k_xa/2$ and $y=\sqrt{3}k_ya/2$.
The TRS operator is given by $\Theta=i\sigma_0s_y\mathsf{K}$ with the complex conjugate operator $\mathsf{K}$.  Under the TRS, the Dirac matrices are even as $\Theta\Gamma^a\Theta^{-1}=\Gamma^a$ while their ten commutators are odd as $\Theta\Gamma^{ab}\Theta^{-1}=-\Gamma^{ab}$. Therefore, the TRS requires $R_a(\bm k)=R_a(-\bm k)$ and $R_{ab}(\bm k)=-R_{ab}(-\bm k)$. These conditions are satisfied in the absence of the phonon dynamics while these terms involving circularly polarized phonons violate the above relations, thereby breaking the TRS in the effective Hamiltonian $\mathcal{H}_{\mathrm{eff}}(\bm k)$.

\begin{figure}
\begin{center}
\includegraphics[width=\columnwidth]{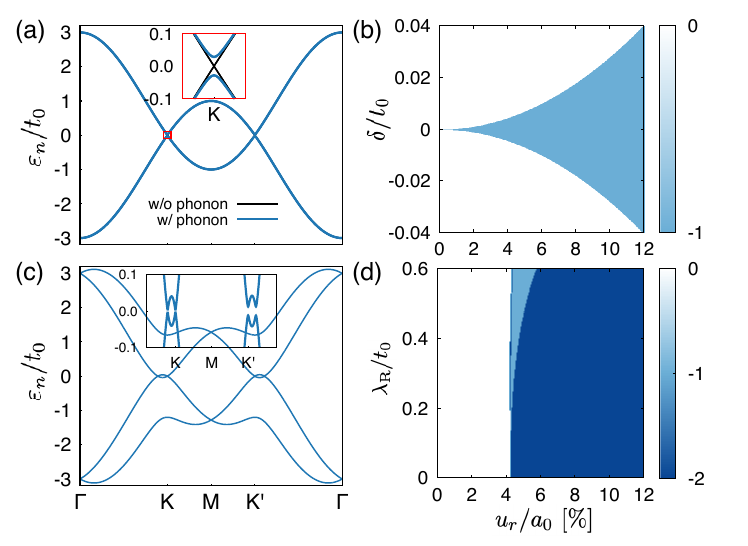}
\caption{Topological phase transition driven by circularly polarized phonons. (a) Band structures of the spinless effective Hamiltonian $h_{\text{eff}}(\bm k)$ along the high-symmetry points with and without phonon dynamics. We set the on-site potential $\delta=0$, and the phonon rotational amplitude $u_r=0.1a_0$, respectively. Inset shows the bands near the $K$ point. (b) Phase diagram of the Chern number as a function of $u_r/a_0$ and $\delta/t_0$. (c) Band structure of the spinful Hamiltonian $\mathcal H_{\text{eff}}(\bm k)$ with the on-site potential $\delta=0.005t_0$, Rashba SOC $\lambda_{\text{R}}=0.4t_0$, and the phonon rotational amplitude $u_r/a_0=5\%$. Inset shows the bands around the valley points. (d) Phase diagram of the Chern number as a function of $u_r/a_0$ and $\lambda_{\text{R}}/t_0$ by fixing the on-site potential as $\delta=0.005t_0$. Here, we set the phonon energy as $\hbar\Omega=0.2t_0$.}
\label{fig2}
\end{center}
\end{figure}

\textit{Topological phase transition.}---
We first consider the case without the Rashba SOC by setting $\lambda_{\text{R}}=0$. The effective Hamiltonian reduces to $\mathcal{H}_{\text{eff}}=h_{\text{eff}}s_0$, where the $2\times2$ spinless Hamiltonian $h_{\text{eff}}=R_1\sigma_x-R_{12}\sigma_y+R_2\sigma_z$ is equivalent to the Haldane model~\cite{Haldane1985}. Within the Floquet picture, the circularly polarized phonons generate an additional mass term in $R_2$ which corresponds to the effective path-dependent NNN hopping from the first term in Eq.~(\ref{H_Omega}). This term further opens a gap and drives a tunable topological phase transition from a trivial insulator to a Chern insulator. We show the band structure with $\delta=0$ in Fig.~\ref{fig2}(a), where a gap opening from an initially massless Dirac dispersion is clearly observed upon switching on the circularly polarized phonon.

We linearize the Hamiltonian as a function of a small wavevector $\bm q=(q_x,q_y)$ around the valley points, which is given by $h_{\text{eff}}^{K/K'}(\bm q)=\mp v_{\text{F}}q_x\sigma_x-v_{\text{F}}q_y\sigma_y+\Delta_{\mp}\sigma_z$ with $v_{\text{F}}=3a_0t_0/2$ being the Fermi velocity divided by $\hbar$, and $\Delta_{\pm}\equiv\delta\pm\frac{3\sqrt{3}\tilde A}{4\hbar\Omega}$ being the gap at the valley points.
Here the phonon-induced gap is given by $2\Delta_{\text{ph}}$, where $\Delta_{\text{ph}}\equiv\frac{3\sqrt{3}\tilde A}{4\hbar\Omega}$. For graphene, the hopping parameter is $t_0=2.7$~eV, and the optical phonon of the in-plane $E_{2g}$ mode has an energy of $\hbar\Omega=0.2$~eV in the terahertz regime~\cite{Yan2007}. For a phonon rotational amplitude of $u_r/a_0=5\%$, the resulting phonon-induced gap is estimated to be around 40meV, which can be observed via angle-resolved photoemission spectroscopy~\cite{Wang2026}.
The Chern number of the whole system is determined by the sign of these masses at the valley points: $C=\frac{1}{2}\left[\text{sgn}(\Delta_-)-\text{sgn}(\Delta_+)\right]$, and phase boundary between topologically trivial and non-trivial phases is given by $\delta=\pm\frac{3\sqrt{3}\tilde A}{4\hbar\Omega}$. The phase diagram in Fig~\ref{fig2}(b) shows the Chern number as functions of $u_r/a_0$ and $\delta/t_0$ for the spinless case.

In the spinful case, when the Rashba SOC is absent, the Chern number becomes $C=-2$ by increasing the phonon rotational amplitude due to the double spin degeneracy. When the Rashba SOC is switched on $(\lambda_{\text{R}}\neq0)$, the spin degeneracy is lifted, and the spin splitting leads to a gap closing again with varying $\lambda_{\text{R}}$. We show the spinful band structure near the topological phase transition from $C=-2$ to $C=-1$ in Fig.~\ref{fig2}(c), where the gap closes and reopens near the $K$ point by varying $\lambda_{\text{R}}$. The phase diagram of the Chern number as a function of $u_r/a_0$ and $\lambda_{\text{R}}/t_0$ with the on-site potential $\delta=0.005t_0$ is shown in Fig.~\ref{fig2}(d). The phonon dynamics with the Rashba SOC leads to topological phase transitions between $C=-2$ and $C=0$~\cite{SM}.
These topological phase transitions can be observed via the Hall-conductance measurement\cite{Kitagawa2011}.

\begin{figure}
\begin{center}
\includegraphics[width=\columnwidth]{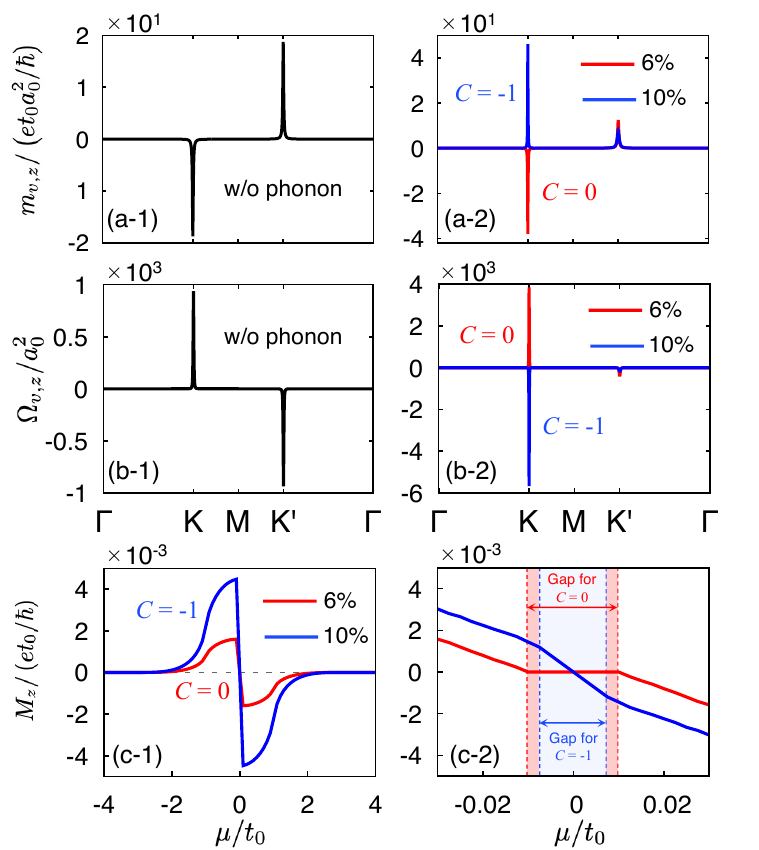}
\caption{Orbital magnetization driven by circularly polarized phonons. (a-1) and (a-2) Intrinsic magnetic moment distribution for the valence band along the high-symmetry points. (b-1) and (b-2) Berry curvature distribution for the valence band along the high-symmetry points. The results without phonon dynamics are shown in (a-1) and (b-1), and those with phonon dynamics for different Chern numbers are shown in (a-2) and (b-2). Red and blue lines in (a-2) and (b-2) represent $u_r/a_0=6\%$ $(C=0)$ and $u_r/a_0=10\%$ $(C=-1)$, respectively. (c-1) and (c-2) Chemical potential dependence of the $z$ component of the orbital magnetization. (c-2) enlarges the region around the band gap. Phonon energy and on-site potential are given by $\hbar\Omega=0.2t_0$ and $\delta=0.02t_0$, respectively.}
\label{fig3}
\end{center}
\end{figure}

\textit{Orbital magnetization.}---
From the modern theory in a semiclassical picture, the orbital magnetization of our electron system at zero temperature is given by
\begin{align}\label{semi_M}
\bm M=\sum_{n}\int_{\varepsilon_{n\bm k}<\mu}\frac{d^2k}{(2\pi)^2}\left\{\bm m_n(\bm k)+\frac{e\bm\Omega_n(\bm k)}{\hbar}[\mu-\varepsilon_{n\bm k}]\right\},
\end{align}
where $\bm m_n(\bm k)=(e/2\hbar)\text{Im}\bra{\partial_{\bm k}u_{n\bm k}}\times[h_{\text{eff}}(\bm k)-\varepsilon_{n\bm k}]\ket{\partial_{\bm k}u_{n\bm k}}$ represents the intrinsic magnetic moment from the self rotation of the electron wave packet with the Bloch state $\ket{u_{n\bm k}}$ and the energy $\varepsilon_{n\bm k}$ of the spinless Hamiltonian $h_{\text{eff}}(\bm k)$, $\bm\Omega_n(\bm k)=i\bra{\partial_{\bm k}u_{n\bm k}}\times\ket{\partial_{\bm k}u_{n\bm k}}$ stands for the Berry curvature, and $\mu$ denotes the chemical potential~\cite{Xiao2005,Thonhauser2005}.
In the absence of phonons, because the TRS is preserved, both $m_{v,z}(\bm k)$ and $\Omega_{v,z}(\bm k)$ for the valence band at the $K$ and $K'$ points exhibit equal magnitudes but opposite signs as shown in Figs.~\ref{fig3}(a-1) and~\ref{fig3}(b-1). As a result, the orbital magnetizations from the two valleys cancel upon integration over the Brillouin zone.

Circularly polarized phonons break the TRS, thereby inducing a nonzero orbital magnetization. In this case, the intrinsic magnetic moment $m_{v,z}(\bm k)$ and Berry curvature $\Omega_{v,z}(\bm k)$ between the two valleys are lifted as shown in Figs~\ref{fig3}(a-2) and~\ref{fig3}(b-2). By using the linearized effective Hamiltonian $h_{\text{eff}}^{K/K'}(\bm q)$, we can express the intrinsic magnetic moment around the valley points as $m_{v,z}^{K/K'}(q)=\mp\frac{e}{2\hbar}\frac{\Delta_{\mp}v_{\text{F}}^2}{v_{\text{F}}^2q^2+\Delta_{\mp}^2}$, and the Berry curvature can be obtained as $\Omega_{v,z}^{K/K'}(q)=\pm\frac{1}{2}\frac{\Delta_{\mp}v_{\text{F}}^2}{(v_{\text{F}}^2q^2+\Delta_{\mp}^2)^{\frac{3}{2}}}$.
The signs of $m^{K/K'}_{v,z}(q)$ and $\Omega^{K/K'}_{v,z}(q)$ around the two valleys are determined by the gap $\Delta_{\pm}$ at the valley points. As the phonon rotational amplitude increases, their signs change, originating from the sign change of $\Delta_{\pm}$ across the topological phase transition.

The orbital magnetization $M_z$ as a function of chemical potential $\mu$ calculated from Eq.~(\ref{semi_M}) is shown in Fig.~\ref{fig3}(c-1), where the red and blue lines denote the rotation amplitudes $u_r/a_0=6\%$ and $u_r/a_0=10\%$ with the Chern numbers $C=0$ and $C=-1$, respectively. Since the intrinsic magnetic moment and Berry curvature of the valence band is connected by $m_{v,z}(\bm k)=(e/2\hbar)[\varepsilon_{v}(\bm k)-\varepsilon_c(\bm k)]\Omega_{v,z}(\bm k)$ with the energy of valence (conduction) band $\varepsilon_{v(c)}(\bm k)$, the $z$ component of the orbital magnetization is expressed as $M_z=(e/2\pi\hbar)\mu C$ with the Chern number $C=\int_{\text{BZ}}\frac{d^2k}{2\pi}\Omega_{v,z}(\bm k)$ when the chemical potential $\mu$ lies within the band gap.
The orbital magnetization for the chemical potential near $\mu=0$ is shown in Fig.~\ref{fig3}(c-2) where the cyan and pink regimes represent the gaps for $C=-1$ and $C=0$, respectively. We notice that the orbital magnetization totally vanishes within the gap when $C=0$, while it shows a linear dependence on the chemical potential $\mu$ inside the gap when $C=-1$.

From the numerical results in Fig.~\ref{fig3}(c-1), the orbital magnetization density reaches $M_z\sim10^{-3}\times et_0/\hbar$. For graphene, the hopping amplitude is $t_0=2.7$ eV, and the orbital magnetization per unit cell is given by $M_{\text{uc}}=M_zS_{\text{uc}}$, where $S_{\text{uc}}=3\sqrt{3}a_0^2/2$ is the unit cell area with the lattice constant $a_0=1.42\mathrm{\AA}$. Thus, the phonon-induced orbital magnetization in graphene is estimated to be $10^{-3}\mu_B$ per unit cell.

\textit{Spin magnetization.}---
Since electron spins do not couple directly to lattice motion, the phonon-driven spin magnetization requires SOC. We next consider the spinful case by switching on the Rashba SOC $(\lambda_{\text{R}}\neq0)$.
The expectation value of electron spin is given by
\begin{align}\label{exp_S}
\braket{\bm S}=\int\frac{d^2k}{(2\pi)^2}\sum_n\langle\hat{\bm S}\rangle_{n\bm k} f(\varepsilon_{n\bm k}),
\end{align}
where $\langle\hat{\bm S}\rangle_{n\bm k}\equiv\langle u_{n\bm k}|\hat{\bm S}|u_{n\bm k}\rangle$ stands for the spin texture of the $n$th state in momentum space~\cite{SM}.
In our system, since the lattice holds the threefold rotation symmetry $C_{3z}$ with respect to the $z$ axis, the in-plane spin magnetization is always zero. It can be easily understood from the real-space Hamiltonian in Eq.~(\ref{H_Omega}), which is invariant under the $C_{3z}$ rotation: $-\cos\frac{\phi_{ij}}{2}s_x+\sin\frac{\phi_{ij}}{2}s_y\rightarrow-\cos\frac{\phi_{ij}}{2}s'_x+\sin\frac{\phi_{ij}}{2}s'_y$ together with the spin rotation $(s'_x,s'_y)=(-\frac{1}{2}s_x-\frac{\sqrt{3}}{2}s_y,\frac{\sqrt{3}}{2}s_x-\frac{1}{2}s_y)$ because the azimuthal angle changes to $\theta_{ki}\rightarrow\theta_{ki}+\frac{2\pi}{3}$ ($\frac{\phi_{ij}}{2}\rightarrow\frac{\phi_{ij}}{2}+\frac{2\pi}{3}$) after $C_{3z}$ rotation.
The spin texture for the $n$th band at each $\bm k$ point satisfies $\braket{\hat{\bm S}}_{n\bm k}=\hat{C}_{3z}^{-1}\braket{\hat{\bm S}}_{n\bm k'}$ with $\bm k'=\hat{C}_{3z}\bm k$. Therefore, the in-plane spin expectation value vanishes after integration over the Brillouin zone.

\begin{figure}
\begin{center}
\includegraphics[width=\columnwidth]{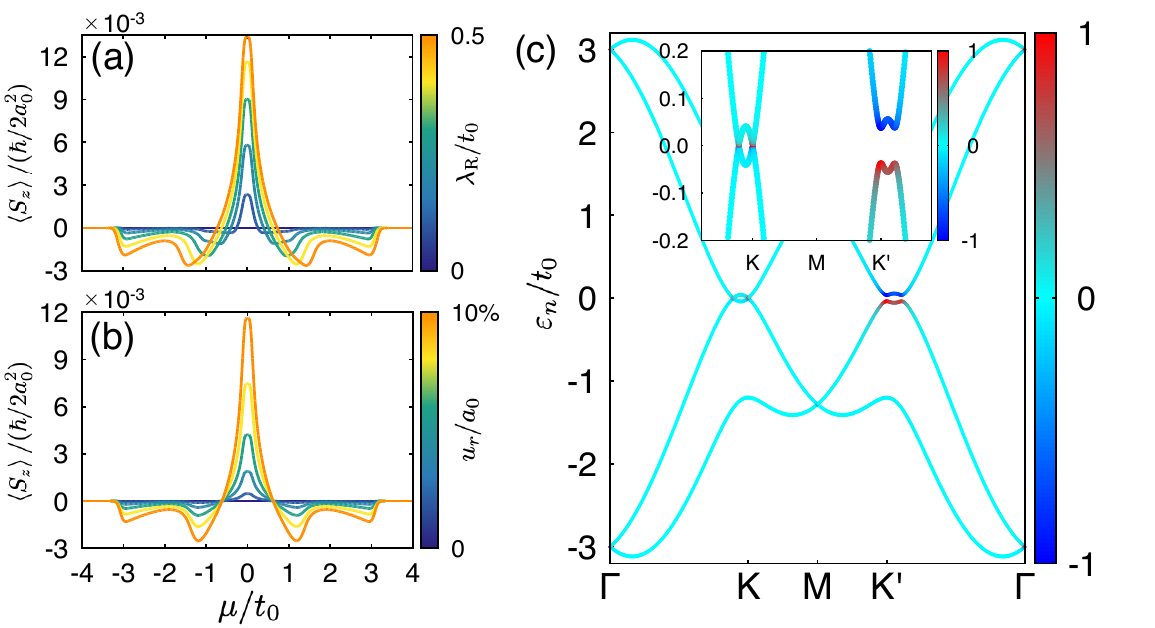}
\caption{Spin magnetization driven by circularly polarized phonons. (a) and (b) Chemical potential dependence of the $z$-component spin expectation value with the colors representing the Rashba SOC and phonon rotational amplitude, respectively. The on-site potential is set to be $\delta=0.02t_0$. We fix the phonon rotational amplitude $u_r/a_0=10\%$ in (a) and the Rashba SOC $\lambda_{\text{R}}=0.4t_0$ in (b). (c) Band structure with the on-site potential $\delta=0.02t_0$, the phonon rotational amplitude $u_r/a_0=8\%$, and the Rashba SOC $\lambda_{\text{R}}=0.4t_0$. The color represents the spin texture $\langle \hat{S}_z\rangle_{n\bm k}/(\hbar/2)$ of each band at $\bm k$ point. Phonon energy and temperature are set to be $\hbar\Omega=0.2t_0$ and $k_{\text{B}}T=0.03t_0$, respectively.}
\label{fig4}
\end{center}
\end{figure}

We then compute the expectation value of the out-of-plane spin magnetization $\braket{S_z}$ by using Eq.~(\ref{exp_S}), where the $z$-component spin operator is replaced by $\hat{S}_z=\frac{\hbar}{2}\sigma_0s_z=\frac{\hbar}{2}\Gamma^{34}$ in our model.
Figures~\ref{fig4}(a) and~\ref{fig4}(b) show the chemical potential dependences of the out-of-plane spin $\braket{S_z}$ with varying the Rashba SOC and the phonon rotational amplitude, respectively. We find that the peak appears at $\mu=0$, because it is dominated by the spin texture $\langle\hat{S}_z\rangle_{n\bm k}$ around the $K'$ point, as indicated by the color scale in Fig.~\ref{fig4}(c).
Meanwhile, the magnitude of $\braket{S_z}$ is enhanced with the increasing Rashba SOC and phonon rotational amplitude.
Our numerical results show that the induced spin magnetization reaches the order of $10^{-3}\mu_B$ per unit cell with the Rashba SOC. Even though the intrinsic SOC in graphene is weak, such spin magnetization can be realized in systems with enhanced interfacial Rashba SOC.

\textit{Conclusion and discussion.---}
In this Letter, we clarify the emergence of magnetism and topology driven by circularly polarized phonons in electron systems via Floquet engineering. By considering a microscopic electron-phonon coupling on a honeycomb lattice, we show that the circularly polarized phonon generates an effective NNN electron hopping, yielding a Haldane-type mass term. This process drives a topological phase transition from a trivial insulator to a Chern insulator, where the Chern number further changes due to the SOC-mediated phonon dynamics.
We also show that orbital and spin magnetizations emerge due to the breaking of TRS.
Our findings show that circularly polarized phonons serve as an effective magnetic field to engineer magnetism and topology, opening a new avenue for phonon Floquet engineering.

At present, we consider the counterclockwise phonon mode at $\Gamma$ point as an example. The clockwise mode is mutually connected via TRS. In this case of the clockwise mode, its Fourier modes $H_{\pm1}$ in the time-dependent Hamiltonian changes to be their Hermitian conjugates, resulting in the opposite sign of Eq.~(\ref{H_Omega}). Accordingly, the signs of the Chern number, orbital and spin magnetizations become opposite.
In nonmagnetic systems, the clockwise and counterclockwise modes are degenerate at the $\Gamma$ point due to TRS. Our proposed effect requires a phonon pumping with terahertz pulses, which leads to an asymmetric population of phonons between the degenerate modes~\cite{Nova2017}.

\begin{acknowledgments}
\textit{Acknowledgments.---}
D.Y. acknowledges fruitful discussions with Professor Shuichi Murakami. D.Y. was supported by Japan Society for the Promotion of Science (JSPS) KAKENHI Grant No.~JP25K23366, and RIKEN Special Postdoctoral Researchers Program.
T.Z. acknowledges from National Key R\&D Project (Grant Nos.~2023YFA1407400 and 2024YFA1409200), and the National Natural Science Foundation of China (Grant Nos.~12374165 and 12447101).
T.O. acknowledges support from JSPS KAKENHI (Nos.~JP23H04865, 23K22487, and 26K00634), MEXT, Japan. 
T.Y. was supported by JSPS KAKENHI No.~JP25K07221.
\end{acknowledgments}

%\bibliography{Floquet.bib}

\begin{thebibliography}{70}%
\makeatletter
\providecommand \@ifxundefined [1]{%
 \@ifx{#1\undefined}
}%
\providecommand \@ifnum [1]{%
 \ifnum #1\expandafter \@firstoftwo
 \else \expandafter \@secondoftwo
 \fi
}%
\providecommand \@ifx [1]{%
 \ifx #1\expandafter \@firstoftwo
 \else \expandafter \@secondoftwo
 \fi
}%
\providecommand \natexlab [1]{#1}%
\providecommand \enquote  [1]{``#1''}%
\providecommand \bibnamefont  [1]{#1}%
\providecommand \bibfnamefont [1]{#1}%
\providecommand \citenamefont [1]{#1}%
\providecommand \href@noop [0]{\@secondoftwo}%
\providecommand \href [0]{\begingroup \@sanitize@url \@href}%
\providecommand \@href[1]{\@@startlink{#1}\@@href}%
\providecommand \@@href[1]{\endgroup#1\@@endlink}%
\providecommand \@sanitize@url [0]{\catcode `\\12\catcode `\$12\catcode
  `\&12\catcode `\#12\catcode `\^12\catcode `\_12\catcode `\%12\relax}%
\providecommand \@@startlink[1]{}%
\providecommand \@@endlink[0]{}%
\providecommand \url  [0]{\begingroup\@sanitize@url \@url }%
\providecommand \@url [1]{\endgroup\@href {#1}{\urlprefix }}%
\providecommand \urlprefix  [0]{URL }%
\providecommand \Eprint [0]{\href }%
\providecommand \doibase [0]{https://doi.org/}%
\providecommand \selectlanguage [0]{\@gobble}%
\providecommand \bibinfo  [0]{\@secondoftwo}%
\providecommand \bibfield  [0]{\@secondoftwo}%
\providecommand \translation [1]{[#1]}%
\providecommand \BibitemOpen [0]{}%
\providecommand \bibitemStop [0]{}%
\providecommand \bibitemNoStop [0]{.\EOS\space}%
\providecommand \EOS [0]{\spacefactor3000\relax}%
\providecommand \BibitemShut  [1]{\csname bibitem#1\endcsname}%
\let\auto@bib@innerbib\@empty
%</preamble>
\bibitem [{\citenamefont {Zhang}\ and\ \citenamefont {Niu}(2014)}]{Zhang2014}%
  \BibitemOpen
  \bibfield  {author} {\bibinfo {author} {\bibfnamefont {L.}~\bibnamefont
  {Zhang}}\ and\ \bibinfo {author} {\bibfnamefont {Q.}~\bibnamefont {Niu}},\
  }\bibfield  {title} {\bibinfo {title} {Angular momentum of phonons and the
  Einstein--de Haas effect},\ }\href
  {https://doi.org/10.1103/PhysRevLett.112.085503} {\bibfield  {journal}
  {\bibinfo  {journal} {Phys. Rev. Lett.}\ }\textbf {\bibinfo {volume} {112}},\
  \bibinfo {pages} {085503} (\bibinfo {year} {2014})}\BibitemShut {NoStop}%
\bibitem [{\citenamefont {Zhang}\ and\ \citenamefont {Niu}(2015)}]{Zhang2015}%
  \BibitemOpen
  \bibfield  {author} {\bibinfo {author} {\bibfnamefont {L.}~\bibnamefont
  {Zhang}}\ and\ \bibinfo {author} {\bibfnamefont {Q.}~\bibnamefont {Niu}},\
  }\bibfield  {title} {\bibinfo {title} {Chiral phonons at high-symmetry points
  in monolayer hexagonal lattices},\ }\href
  {https://doi.org/10.1103/PhysRevLett.115.115502} {\bibfield  {journal}
  {\bibinfo  {journal} {Phys. Rev. Lett.}\ }\textbf {\bibinfo {volume} {115}},\
  \bibinfo {pages} {115502} (\bibinfo {year} {2015})}\BibitemShut {NoStop}%
\bibitem [{\citenamefont {Zhu}\ \emph {et~al.}(2018)\citenamefont {Zhu},
  \citenamefont {Yi}, \citenamefont {Li}, \citenamefont {Xiao}, \citenamefont
  {Zhang}, \citenamefont {Yang}, \citenamefont {Kaindl}, \citenamefont {Li},
  \citenamefont {Wang},\ and\ \citenamefont {Zhang}}]{Zhu2018}%
  \BibitemOpen
  \bibfield  {author} {\bibinfo {author} {\bibfnamefont {H.}~\bibnamefont
  {Zhu}}, \bibinfo {author} {\bibfnamefont {J.}~\bibnamefont {Yi}}, \bibinfo
  {author} {\bibfnamefont {M.-Y.}\ \bibnamefont {Li}}, \bibinfo {author}
  {\bibfnamefont {J.}~\bibnamefont {Xiao}}, \bibinfo {author} {\bibfnamefont
  {L.}~\bibnamefont {Zhang}}, \bibinfo {author} {\bibfnamefont {C.-W.}\
  \bibnamefont {Yang}}, \bibinfo {author} {\bibfnamefont {R.~A.}\ \bibnamefont
  {Kaindl}}, \bibinfo {author} {\bibfnamefont {L.-J.}\ \bibnamefont {Li}},
  \bibinfo {author} {\bibfnamefont {Y.}~\bibnamefont {Wang}},\ and\ \bibinfo
  {author} {\bibfnamefont {X.}~\bibnamefont {Zhang}},\ }\bibfield  {title}
  {\bibinfo {title} {Observation of chiral phonons},\ }\href
  {https://doi.org/10.1126/science.aar2711} {\bibfield  {journal} {\bibinfo
  {journal} {Science}\ }\textbf {\bibinfo {volume} {359}},\ \bibinfo {pages}
  {579} (\bibinfo {year} {2018})}\BibitemShut {NoStop}%
\bibitem [{\citenamefont {Juraschek}\ and\ \citenamefont
  {Spaldin}(2019)}]{Juraschek2019}%
  \BibitemOpen
  \bibfield  {author} {\bibinfo {author} {\bibfnamefont {D.~M.}\ \bibnamefont
  {Juraschek}}\ and\ \bibinfo {author} {\bibfnamefont {N.~A.}\ \bibnamefont
  {Spaldin}},\ }\bibfield  {title} {\bibinfo {title} {Orbital magnetic moments
  of phonons},\ }\href {https://doi.org/10.1103/PhysRevMaterials.3.064405}
  {\bibfield  {journal} {\bibinfo  {journal} {Phys. Rev. Mater.}\ }\textbf
  {\bibinfo {volume} {3}},\ \bibinfo {pages} {064405} (\bibinfo {year}
  {2019})}\BibitemShut {NoStop}%
\bibitem [{\citenamefont {Zhang}\ and\ \citenamefont
  {Murakami}(2022)}]{Zhang2022}%
  \BibitemOpen
  \bibfield  {author} {\bibinfo {author} {\bibfnamefont {T.}~\bibnamefont
  {Zhang}}\ and\ \bibinfo {author} {\bibfnamefont {S.}~\bibnamefont
  {Murakami}},\ }\bibfield  {title} {\bibinfo {title} {Chiral phonons and
  pseudoangular momentum in nonsymmorphic systems},\ }\href
  {https://doi.org/10.1103/PhysRevResearch.4.L012024} {\bibfield  {journal}
  {\bibinfo  {journal} {Phys. Rev. Res.}\ }\textbf {\bibinfo {volume} {4}},\
  \bibinfo {pages} {L012024} (\bibinfo {year} {2022})}\BibitemShut {NoStop}%
\bibitem [{\citenamefont {Zhang}\ \emph {et~al.}(2023)\citenamefont {Zhang},
  \citenamefont {Huang}, \citenamefont {Pan}, \citenamefont {Du}, \citenamefont
  {Zhang},\ and\ \citenamefont {Murakami}}]{Zhang2023}%
  \BibitemOpen
  \bibfield  {author} {\bibinfo {author} {\bibfnamefont {T.}~\bibnamefont
  {Zhang}}, \bibinfo {author} {\bibfnamefont {Z.}~\bibnamefont {Huang}},
  \bibinfo {author} {\bibfnamefont {Z.}~\bibnamefont {Pan}}, \bibinfo {author}
  {\bibfnamefont {L.}~\bibnamefont {Du}}, \bibinfo {author} {\bibfnamefont
  {G.}~\bibnamefont {Zhang}},\ and\ \bibinfo {author} {\bibfnamefont
  {S.}~\bibnamefont {Murakami}},\ }\bibfield  {title} {\bibinfo {title} {Weyl
  phonons in chiral crystals},\ }\href
  {https://doi.org/10.1021/acs.nanolett.3c02132} {\bibfield  {journal}
  {\bibinfo  {journal} {Nano Letters}\ }\textbf {\bibinfo {volume} {23}},\
  \bibinfo {pages} {7561} (\bibinfo {year} {2023})}\BibitemShut {NoStop}%
\bibitem [{\citenamefont {Ishito}\ \emph {et~al.}(2023)\citenamefont {Ishito},
  \citenamefont {Mao}, \citenamefont {Kousaka}, \citenamefont {Togawa},
  \citenamefont {Iwasaki}, \citenamefont {Zhang}, \citenamefont {Murakami},
  \citenamefont {Kishine},\ and\ \citenamefont {Satoh}}]{Ishito2023}%
  \BibitemOpen
  \bibfield  {author} {\bibinfo {author} {\bibfnamefont {K.}~\bibnamefont
  {Ishito}}, \bibinfo {author} {\bibfnamefont {H.}~\bibnamefont {Mao}},
  \bibinfo {author} {\bibfnamefont {Y.}~\bibnamefont {Kousaka}}, \bibinfo
  {author} {\bibfnamefont {Y.}~\bibnamefont {Togawa}}, \bibinfo {author}
  {\bibfnamefont {S.}~\bibnamefont {Iwasaki}}, \bibinfo {author} {\bibfnamefont
  {T.}~\bibnamefont {Zhang}}, \bibinfo {author} {\bibfnamefont
  {S.}~\bibnamefont {Murakami}}, \bibinfo {author} {\bibfnamefont {J.-i.}\
  \bibnamefont {Kishine}},\ and\ \bibinfo {author} {\bibfnamefont
  {T.}~\bibnamefont {Satoh}},\ }\bibfield  {title} {\bibinfo {title} {Truly
  chiral phonons in $\alpha$-HgS},\ }\href
  {https://doi.org/10.1038/s41567-022-01790-x} {\bibfield  {journal} {\bibinfo
  {journal} {Nature Physics}\ }\textbf {\bibinfo {volume} {19}},\ \bibinfo
  {pages} {35} (\bibinfo {year} {2023})}\BibitemShut {NoStop}%
\bibitem [{\citenamefont {Ueda}\ \emph {et~al.}(2023)\citenamefont {Ueda},
  \citenamefont {Garc{\'i}a-Fern{\'a}ndez}, \citenamefont {Agrestini},
  \citenamefont {Romao}, \citenamefont {van~den Brink}, \citenamefont
  {Spaldin}, \citenamefont {Zhou},\ and\ \citenamefont {Staub}}]{Ueda2023}%
  \BibitemOpen
  \bibfield  {author} {\bibinfo {author} {\bibfnamefont {H.}~\bibnamefont
  {Ueda}}, \bibinfo {author} {\bibfnamefont {M.}~\bibnamefont
  {Garc{\'i}a-Fern{\'a}ndez}}, \bibinfo {author} {\bibfnamefont
  {S.}~\bibnamefont {Agrestini}}, \bibinfo {author} {\bibfnamefont {C.~P.}\
  \bibnamefont {Romao}}, \bibinfo {author} {\bibfnamefont {J.}~\bibnamefont
  {van~den Brink}}, \bibinfo {author} {\bibfnamefont {N.~A.}\ \bibnamefont
  {Spaldin}}, \bibinfo {author} {\bibfnamefont {K.-J.}\ \bibnamefont {Zhou}},\
  and\ \bibinfo {author} {\bibfnamefont {U.}~\bibnamefont {Staub}},\ }\bibfield
   {title} {\bibinfo {title} {Chiral phonons in quartz probed by X-rays},\
  }\href {https://doi.org/10.1038/s41586-023-06016-5} {\bibfield  {journal}
  {\bibinfo  {journal} {Nature}\ }\textbf {\bibinfo {volume} {618}},\ \bibinfo
  {pages} {946} (\bibinfo {year} {2023})}\BibitemShut {NoStop}%
\bibitem [{\citenamefont {Ohe}\ \emph {et~al.}(2024)\citenamefont {Ohe},
  \citenamefont {Shishido}, \citenamefont {Kato}, \citenamefont {Utsumi},
  \citenamefont {Matsuura},\ and\ \citenamefont {Togawa}}]{Ohe2024}%
  \BibitemOpen
  \bibfield  {author} {\bibinfo {author} {\bibfnamefont {K.}~\bibnamefont
  {Ohe}}, \bibinfo {author} {\bibfnamefont {H.}~\bibnamefont {Shishido}},
  \bibinfo {author} {\bibfnamefont {M.}~\bibnamefont {Kato}}, \bibinfo {author}
  {\bibfnamefont {S.}~\bibnamefont {Utsumi}}, \bibinfo {author} {\bibfnamefont
  {H.}~\bibnamefont {Matsuura}},\ and\ \bibinfo {author} {\bibfnamefont
  {Y.}~\bibnamefont {Togawa}},\ }\bibfield  {title} {\bibinfo {title}
  {Chirality-induced selectivity of phonon angular momenta in chiral quartz
  crystals},\ }\href {https://doi.org/10.1103/PhysRevLett.132.056302}
  {\bibfield  {journal} {\bibinfo  {journal} {Phys. Rev. Lett.}\ }\textbf
  {\bibinfo {volume} {132}},\ \bibinfo {pages} {056302} (\bibinfo {year}
  {2024})}\BibitemShut {NoStop}%
\bibitem [{\citenamefont {Zhang}\ \emph
  {et~al.}(2025{\natexlab{a}})\citenamefont {Zhang}, \citenamefont {Liu},
  \citenamefont {Miao},\ and\ \citenamefont {Murakami}}]{Zhang2025}%
  \BibitemOpen
  \bibfield  {author} {\bibinfo {author} {\bibfnamefont {T.}~\bibnamefont
  {Zhang}}, \bibinfo {author} {\bibfnamefont {Y.}~\bibnamefont {Liu}}, \bibinfo
  {author} {\bibfnamefont {H.}~\bibnamefont {Miao}},\ and\ \bibinfo {author}
  {\bibfnamefont {S.}~\bibnamefont {Murakami}},\ }\href@noop {} {\bibinfo
  {title} {New advances in phonons: From band topology to quasiparticle
  chirality}} (\bibinfo {year} {2025}{\natexlab{a}}),\ \Eprint
  {https://arxiv.org/abs/arXiv:2505.06179} {arXiv:2505.06179} \BibitemShut
  {NoStop}%
\bibitem [{\citenamefont {Zhang}\ \emph
  {et~al.}(2025{\natexlab{b}})\citenamefont {Zhang}, \citenamefont {Murakami},\
  and\ \citenamefont {Miao}}]{Zhang2025NC}%
  \BibitemOpen
  \bibfield  {author} {\bibinfo {author} {\bibfnamefont {T.}~\bibnamefont
  {Zhang}}, \bibinfo {author} {\bibfnamefont {S.}~\bibnamefont {Murakami}},\
  and\ \bibinfo {author} {\bibfnamefont {H.}~\bibnamefont {Miao}},\ }\bibfield
  {title} {\bibinfo {title} {Weyl phonons: the connection of topology and
  chirality},\ }\href {https://doi.org/10.1038/s41467-025-58913-0} {\bibfield
  {journal} {\bibinfo  {journal} {Nature Communications}\ }\textbf {\bibinfo
  {volume} {16}},\ \bibinfo {pages} {3560} (\bibinfo {year}
  {2025}{\natexlab{b}})}\BibitemShut {NoStop}%
\bibitem [{\citenamefont {Juraschek}\ \emph {et~al.}(2025)\citenamefont
  {Juraschek}, \citenamefont {Geilhufe}, \citenamefont {Zhu}, \citenamefont
  {Basini}, \citenamefont {Baum}, \citenamefont {Baydin}, \citenamefont
  {Chaudhary}, \citenamefont {Fechner}, \citenamefont {Flebus}, \citenamefont
  {Grissonnanche}, \citenamefont {Kirilyuk}, \citenamefont {Lemeshko},
  \citenamefont {Maehrlein}, \citenamefont {Mignolet}, \citenamefont
  {Murakami}, \citenamefont {Niu}, \citenamefont {Nowak}, \citenamefont
  {Romao}, \citenamefont {Rostami}, \citenamefont {Satoh}, \citenamefont
  {Spaldin}, \citenamefont {Ueda},\ and\ \citenamefont
  {Zhang}}]{Juraschek2025}%
  \BibitemOpen
  \bibfield  {author} {\bibinfo {author} {\bibfnamefont {D.~M.}\ \bibnamefont
  {Juraschek}}, \bibinfo {author} {\bibfnamefont {R.~M.}\ \bibnamefont
  {Geilhufe}}, \bibinfo {author} {\bibfnamefont {H.}~\bibnamefont {Zhu}},
  \bibinfo {author} {\bibfnamefont {M.}~\bibnamefont {Basini}}, \bibinfo
  {author} {\bibfnamefont {P.}~\bibnamefont {Baum}}, \bibinfo {author}
  {\bibfnamefont {A.}~\bibnamefont {Baydin}}, \bibinfo {author} {\bibfnamefont
  {S.}~\bibnamefont {Chaudhary}}, \bibinfo {author} {\bibfnamefont
  {M.}~\bibnamefont {Fechner}}, \bibinfo {author} {\bibfnamefont
  {B.}~\bibnamefont {Flebus}}, \bibinfo {author} {\bibfnamefont
  {G.}~\bibnamefont {Grissonnanche}}, \bibinfo {author} {\bibfnamefont {A.~I.}\
  \bibnamefont {Kirilyuk}}, \bibinfo {author} {\bibfnamefont {M.}~\bibnamefont
  {Lemeshko}}, \bibinfo {author} {\bibfnamefont {S.~F.}\ \bibnamefont
  {Maehrlein}}, \bibinfo {author} {\bibfnamefont {M.}~\bibnamefont {Mignolet}},
  \bibinfo {author} {\bibfnamefont {S.}~\bibnamefont {Murakami}}, \bibinfo
  {author} {\bibfnamefont {Q.}~\bibnamefont {Niu}}, \bibinfo {author}
  {\bibfnamefont {U.}~\bibnamefont {Nowak}}, \bibinfo {author} {\bibfnamefont
  {C.~P.}\ \bibnamefont {Romao}}, \bibinfo {author} {\bibfnamefont
  {H.}~\bibnamefont {Rostami}}, \bibinfo {author} {\bibfnamefont
  {T.}~\bibnamefont {Satoh}}, \bibinfo {author} {\bibfnamefont {N.~A.}\
  \bibnamefont {Spaldin}}, \bibinfo {author} {\bibfnamefont {H.}~\bibnamefont
  {Ueda}},\ and\ \bibinfo {author} {\bibfnamefont {L.}~\bibnamefont {Zhang}},\
  }\bibfield  {title} {\bibinfo {title} {Chiral phonons},\ }\href
  {https://doi.org/10.1038/s41567-025-03001-9} {\bibfield  {journal} {\bibinfo
  {journal} {Nature Physics}\ }\textbf {\bibinfo {volume} {21}},\ \bibinfo
  {pages} {1532} (\bibinfo {year} {2025})}\BibitemShut {NoStop}%
\bibitem [{\citenamefont {Shabala}\ \emph {et~al.}(2025)\citenamefont
  {Shabala}, \citenamefont {Tietjen},\ and\ \citenamefont
  {Geilhufe}}]{Natalia2025}%
  \BibitemOpen
  \bibfield  {author} {\bibinfo {author} {\bibfnamefont {N.}~\bibnamefont
  {Shabala}}, \bibinfo {author} {\bibfnamefont {F.}~\bibnamefont {Tietjen}},\
  and\ \bibinfo {author} {\bibfnamefont {R.~M.}\ \bibnamefont {Geilhufe}},\
  }\href@noop {} {\bibinfo {title} {Axial phono-magnetic effects}} (\bibinfo
  {year} {2025}),\ \Eprint {https://arxiv.org/abs/arXiv:2511.03329}
  {arXiv:2511.03329} \BibitemShut {NoStop}%
\bibitem [{\citenamefont {Che}\ \emph {et~al.}(2025)\citenamefont {Che},
  \citenamefont {Liang}, \citenamefont {Cui}, \citenamefont {Li}, \citenamefont
  {Lu}, \citenamefont {Sang}, \citenamefont {Li}, \citenamefont {Dong},
  \citenamefont {Zhao}, \citenamefont {Zhang}, \citenamefont {Sun},
  \citenamefont {Jiang}, \citenamefont {Liu}, \citenamefont {Jin},
  \citenamefont {Zhang},\ and\ \citenamefont {Yang}}]{Che2025}%
  \BibitemOpen
  \bibfield  {author} {\bibinfo {author} {\bibfnamefont {M.}~\bibnamefont
  {Che}}, \bibinfo {author} {\bibfnamefont {J.}~\bibnamefont {Liang}}, \bibinfo
  {author} {\bibfnamefont {Y.}~\bibnamefont {Cui}}, \bibinfo {author}
  {\bibfnamefont {H.}~\bibnamefont {Li}}, \bibinfo {author} {\bibfnamefont
  {B.}~\bibnamefont {Lu}}, \bibinfo {author} {\bibfnamefont {W.}~\bibnamefont
  {Sang}}, \bibinfo {author} {\bibfnamefont {X.}~\bibnamefont {Li}}, \bibinfo
  {author} {\bibfnamefont {X.}~\bibnamefont {Dong}}, \bibinfo {author}
  {\bibfnamefont {L.}~\bibnamefont {Zhao}}, \bibinfo {author} {\bibfnamefont
  {S.}~\bibnamefont {Zhang}}, \bibinfo {author} {\bibfnamefont
  {T.}~\bibnamefont {Sun}}, \bibinfo {author} {\bibfnamefont {W.}~\bibnamefont
  {Jiang}}, \bibinfo {author} {\bibfnamefont {E.}~\bibnamefont {Liu}}, \bibinfo
  {author} {\bibfnamefont {F.}~\bibnamefont {Jin}}, \bibinfo {author}
  {\bibfnamefont {T.}~\bibnamefont {Zhang}},\ and\ \bibinfo {author}
  {\bibfnamefont {L.}~\bibnamefont {Yang}},\ }\bibfield  {title} {\bibinfo
  {title} {Magnetic order induced chiral phonons in a ferromagnetic Weyl
  semimetal},\ }\href {https://doi.org/10.1103/PhysRevLett.134.196906}
  {\bibfield  {journal} {\bibinfo  {journal} {Phys. Rev. Lett.}\ }\textbf
  {\bibinfo {volume} {134}},\ \bibinfo {pages} {196906} (\bibinfo {year}
  {2025})}\BibitemShut {NoStop}%
\bibitem [{\citenamefont {Yang}\ \emph {et~al.}(2025)\citenamefont {Yang},
  \citenamefont {Zhu}, \citenamefont {Steigleder}, \citenamefont {Liu},
  \citenamefont {Liu}, \citenamefont {Qiu}, \citenamefont {Zhang},\ and\
  \citenamefont {Dressel}}]{Yang2025}%
  \BibitemOpen
  \bibfield  {author} {\bibinfo {author} {\bibfnamefont {R.}~\bibnamefont
  {Yang}}, \bibinfo {author} {\bibfnamefont {Y.-Y.}\ \bibnamefont {Zhu}},
  \bibinfo {author} {\bibfnamefont {M.}~\bibnamefont {Steigleder}}, \bibinfo
  {author} {\bibfnamefont {Y.-C.}\ \bibnamefont {Liu}}, \bibinfo {author}
  {\bibfnamefont {C.-C.}\ \bibnamefont {Liu}}, \bibinfo {author} {\bibfnamefont
  {X.-G.}\ \bibnamefont {Qiu}}, \bibinfo {author} {\bibfnamefont
  {T.}~\bibnamefont {Zhang}},\ and\ \bibinfo {author} {\bibfnamefont
  {M.}~\bibnamefont {Dressel}},\ }\bibfield  {title} {\bibinfo {title}
  {Inherent circular dichroism of phonons in magnetic Weyl semimetal
  ${\mathrm{co}}_{3}{\text{sn}}_{2}{\mathrm{s}}_{2}$},\ }\href
  {https://doi.org/10.1103/PhysRevLett.134.196905} {\bibfield  {journal}
  {\bibinfo  {journal} {Phys. Rev. Lett.}\ }\textbf {\bibinfo {volume} {134}},\
  \bibinfo {pages} {196905} (\bibinfo {year} {2025})}\BibitemShut {NoStop}%
\bibitem [{\citenamefont {Zhang}\ \emph {et~al.}(2026)\citenamefont {Zhang},
  \citenamefont {Huang}, \citenamefont {Du}, \citenamefont {Ying},
  \citenamefont {Du},\ and\ \citenamefont {Zhang}}]{S_Zhang2026}%
  \BibitemOpen
  \bibfield  {author} {\bibinfo {author} {\bibfnamefont {S.}~\bibnamefont
  {Zhang}}, \bibinfo {author} {\bibfnamefont {Z.}~\bibnamefont {Huang}},
  \bibinfo {author} {\bibfnamefont {M.}~\bibnamefont {Du}}, \bibinfo {author}
  {\bibfnamefont {T.}~\bibnamefont {Ying}}, \bibinfo {author} {\bibfnamefont
  {L.}~\bibnamefont {Du}},\ and\ \bibinfo {author} {\bibfnamefont
  {T.}~\bibnamefont {Zhang}},\ }\bibfield  {title} {\bibinfo {title}
  {Comprehensive study of phonon chirality under symmetry constraints},\ }\href
  {https://doi.org/10.1103/gmfc-gx4v} {\bibfield  {journal} {\bibinfo
  {journal} {Phys. Rev. B}\ }\textbf {\bibinfo {volume} {113}},\ \bibinfo
  {pages} {024302} (\bibinfo {year} {2026})}\BibitemShut {NoStop}%
\bibitem [{\citenamefont {Juraschek}\ \emph {et~al.}(2020)\citenamefont
  {Juraschek}, \citenamefont {Narang},\ and\ \citenamefont
  {Spaldin}}]{Juraschek2020}%
  \BibitemOpen
  \bibfield  {author} {\bibinfo {author} {\bibfnamefont {D.~M.}\ \bibnamefont
  {Juraschek}}, \bibinfo {author} {\bibfnamefont {P.}~\bibnamefont {Narang}},\
  and\ \bibinfo {author} {\bibfnamefont {N.~A.}\ \bibnamefont {Spaldin}},\
  }\bibfield  {title} {\bibinfo {title} {Phono-magnetic analogs to
  opto-magnetic effects},\ }\href
  {https://doi.org/10.1103/PhysRevResearch.2.043035} {\bibfield  {journal}
  {\bibinfo  {journal} {Phys. Rev. Res.}\ }\textbf {\bibinfo {volume} {2}},\
  \bibinfo {pages} {043035} (\bibinfo {year} {2020})}\BibitemShut {NoStop}%
\bibitem [{\citenamefont {Juraschek}\ \emph {et~al.}(2022)\citenamefont
  {Juraschek}, \citenamefont {Neuman},\ and\ \citenamefont
  {Narang}}]{Juraschek2022}%
  \BibitemOpen
  \bibfield  {author} {\bibinfo {author} {\bibfnamefont {D.~M.}\ \bibnamefont
  {Juraschek}}, \bibinfo {author} {\bibfnamefont {T.}\ \bibnamefont
  {Neuman}},\ and\ \bibinfo {author} {\bibfnamefont {P.}~\bibnamefont
  {Narang}},\ }\bibfield  {title} {\bibinfo {title} {Giant effective magnetic
  fields from optically driven chiral phonons in $4f$ paramagnets},\ }\href
  {https://doi.org/10.1103/PhysRevResearch.4.013129} {\bibfield  {journal}
  {\bibinfo  {journal} {Phys. Rev. Res.}\ }\textbf {\bibinfo {volume} {4}},\
  \bibinfo {pages} {013129} (\bibinfo {year} {2022})}\BibitemShut {NoStop}%
\bibitem [{\citenamefont {Luo}\ \emph {et~al.}(2023)\citenamefont {Luo},
  \citenamefont {Lin}, \citenamefont {Zhang}, \citenamefont {Chen},
  \citenamefont {Blackert}, \citenamefont {Xu}, \citenamefont {Yakobson},\ and\
  \citenamefont {Zhu}}]{Luo2023}%
  \BibitemOpen
  \bibfield  {author} {\bibinfo {author} {\bibfnamefont {J.}~\bibnamefont
  {Luo}}, \bibinfo {author} {\bibfnamefont {T.}~\bibnamefont {Lin}}, \bibinfo
  {author} {\bibfnamefont {J.}~\bibnamefont {Zhang}}, \bibinfo {author}
  {\bibfnamefont {X.}~\bibnamefont {Chen}}, \bibinfo {author} {\bibfnamefont
  {E.~R.}\ \bibnamefont {Blackert}}, \bibinfo {author} {\bibfnamefont
  {R.}~\bibnamefont {Xu}}, \bibinfo {author} {\bibfnamefont {B.~I.}\
  \bibnamefont {Yakobson}},\ and\ \bibinfo {author} {\bibfnamefont
  {H.}~\bibnamefont {Zhu}},\ }\bibfield  {title} {\bibinfo {title} {Large
  effective magnetic fields from chiral phonons in rare-earth halides},\ }\href
  {https://doi.org/10.1126/science.adi9601} {\bibfield  {journal} {\bibinfo
  {journal} {Science}\ }\textbf {\bibinfo {volume} {382}},\ \bibinfo {pages}
  {698} (\bibinfo {year} {2023})}\BibitemShut {NoStop}%
\bibitem [{\citenamefont {Kim}\ \emph {et~al.}(2023)\citenamefont {Kim},
  \citenamefont {Vetter}, \citenamefont {Yan}, \citenamefont {Yang},
  \citenamefont {Wang}, \citenamefont {Sun}, \citenamefont {Yang},
  \citenamefont {Comstock}, \citenamefont {Li}, \citenamefont {Zhou},
  \citenamefont {Zhang}, \citenamefont {You}, \citenamefont {Sun},\ and\
  \citenamefont {Liu}}]{Kim2023}%
  \BibitemOpen
  \bibfield  {author} {\bibinfo {author} {\bibfnamefont {K.}~\bibnamefont
  {Kim}}, \bibinfo {author} {\bibfnamefont {E.}~\bibnamefont {Vetter}},
  \bibinfo {author} {\bibfnamefont {L.}~\bibnamefont {Yan}}, \bibinfo {author}
  {\bibfnamefont {C.}~\bibnamefont {Yang}}, \bibinfo {author} {\bibfnamefont
  {Z.}~\bibnamefont {Wang}}, \bibinfo {author} {\bibfnamefont {R.}~\bibnamefont
  {Sun}}, \bibinfo {author} {\bibfnamefont {Y.}~\bibnamefont {Yang}}, \bibinfo
  {author} {\bibfnamefont {A.~H.}\ \bibnamefont {Comstock}}, \bibinfo {author}
  {\bibfnamefont {X.}~\bibnamefont {Li}}, \bibinfo {author} {\bibfnamefont
  {J.}~\bibnamefont {Zhou}}, \bibinfo {author} {\bibfnamefont {L.}~\bibnamefont
  {Zhang}}, \bibinfo {author} {\bibfnamefont {W.}~\bibnamefont {You}}, \bibinfo
  {author} {\bibfnamefont {D.}~\bibnamefont {Sun}},\ and\ \bibinfo {author}
  {\bibfnamefont {J.}~\bibnamefont {Liu}},\ }\bibfield  {title} {\bibinfo
  {title} {Chiral-phonon-activated spin Seebeck effect},\ }\href
  {https://doi.org/10.1038/s41563-023-01473-9} {\bibfield  {journal} {\bibinfo
  {journal} {Nature Materials}\ }\textbf {\bibinfo {volume} {22}},\ \bibinfo
  {pages} {322} (\bibinfo {year} {2023})}\BibitemShut {NoStop}%
\bibitem [{\citenamefont {Geilhufe}\ and\ \citenamefont
  {Hergert}(2023)}]{Geilhufe2023}%
  \BibitemOpen
  \bibfield  {author} {\bibinfo {author} {\bibfnamefont {R.~M.}\ \bibnamefont
  {Geilhufe}}\ and\ \bibinfo {author} {\bibfnamefont {W.}~\bibnamefont
  {Hergert}},\ }\bibfield  {title} {\bibinfo {title} {Electron magnetic moment
  of transient chiral phonons in ${\mathrm{KTaO}}_{3}$},\ }\href
  {https://doi.org/10.1103/PhysRevB.107.L020406} {\bibfield  {journal}
  {\bibinfo  {journal} {Phys. Rev. B}\ }\textbf {\bibinfo {volume} {107}},\
  \bibinfo {pages} {L020406} (\bibinfo {year} {2023})}\BibitemShut {NoStop}%
\bibitem [{\citenamefont {Yao}\ \emph {et~al.}(2024)\citenamefont {Yao},
  \citenamefont {Matsuo},\ and\ \citenamefont {Yokoyama}}]{YaoAPL2024}%
  \BibitemOpen
  \bibfield  {author} {\bibinfo {author} {\bibfnamefont {D.}~\bibnamefont
  {Yao}}, \bibinfo {author} {\bibfnamefont {M.}~\bibnamefont {Matsuo}},\ and\
  \bibinfo {author} {\bibfnamefont {T.}~\bibnamefont {Yokoyama}},\ }\bibfield
  {title} {\bibinfo {title} {{Electric field-induced nonreciprocal spin current
  due to chiral phonons in chiral-structure superconductors}},\ }\href
  {https://doi.org/10.1063/5.0207915} {\bibfield  {journal} {\bibinfo
  {journal} {Applied Physics Letters}\ }\textbf {\bibinfo {volume} {124}},\
  \bibinfo {pages} {162603} (\bibinfo {year} {2024})}\BibitemShut {NoStop}%
\bibitem [{\citenamefont {Funato}\ \emph {et~al.}(2024)\citenamefont {Funato},
  \citenamefont {Matsuo},\ and\ \citenamefont {Kato}}]{Funato2024}%
  \BibitemOpen
  \bibfield  {author} {\bibinfo {author} {\bibfnamefont {T.}~\bibnamefont
  {Funato}}, \bibinfo {author} {\bibfnamefont {M.}~\bibnamefont {Matsuo}},\
  and\ \bibinfo {author} {\bibfnamefont {T.}~\bibnamefont {Kato}},\ }\bibfield
  {title} {\bibinfo {title} {Chirality-induced phonon-spin conversion at an
  interface},\ }\href {https://doi.org/10.1103/PhysRevLett.132.236201}
  {\bibfield  {journal} {\bibinfo  {journal} {Phys. Rev. Lett.}\ }\textbf
  {\bibinfo {volume} {132}},\ \bibinfo {pages} {236201} (\bibinfo {year}
  {2024})}\BibitemShut {NoStop}%
\bibitem [{\citenamefont {Yokoyama}(2024)}]{Yokoyama2024}%
  \BibitemOpen
  \bibfield  {author} {\bibinfo {author} {\bibfnamefont {T.}~\bibnamefont
  {Yokoyama}},\ }\bibfield  {title} {\bibinfo {title} {Spin-spin interaction
  mediated by chiral phonons},\ }\href {https://doi.org/10.7566/JPSJ.93.123705}
  {\bibfield  {journal} {\bibinfo  {journal} {J. Phys. Soc. Jpn.}\ }\textbf
  {\bibinfo {volume} {93}},\ \bibinfo {pages} {123705} (\bibinfo {year}
  {2024})}\BibitemShut {NoStop}%
\bibitem [{\citenamefont {Shabala}\ and\ \citenamefont
  {Geilhufe}(2024)}]{Shabala2024}%
  \BibitemOpen
  \bibfield  {author} {\bibinfo {author} {\bibfnamefont {N.}~\bibnamefont
  {Shabala}}\ and\ \bibinfo {author} {\bibfnamefont {R.~M.}\ \bibnamefont
  {Geilhufe}},\ }\bibfield  {title} {\bibinfo {title} {Phonon inverse Faraday
  effect from electron-phonon coupling},\ }\href
  {https://doi.org/10.1103/PhysRevLett.133.266702} {\bibfield  {journal}
  {\bibinfo  {journal} {Phys. Rev. Lett.}\ }\textbf {\bibinfo {volume} {133}},\
  \bibinfo {pages} {266702} (\bibinfo {year} {2024})}\BibitemShut {NoStop}%
\bibitem [{\citenamefont {Yokoyama}(2025)}]{Yokoyama2025}%
  \BibitemOpen
  \bibfield  {author} {\bibinfo {author} {\bibfnamefont {T.}~\bibnamefont
  {Yokoyama}},\ }\bibfield  {title} {\bibinfo {title} {Phonon Edelstein effect
  in chiral metals},\ }\href {https://doi.org/10.1103/lsv4-z14s} {\bibfield
  {journal} {\bibinfo  {journal} {Phys. Rev. B}\ }\textbf {\bibinfo {volume}
  {112}},\ \bibinfo {pages} {L020406} (\bibinfo {year} {2025})}\BibitemShut
  {NoStop}%
\bibitem [{\citenamefont {Nishimura}\ \emph {et~al.}(2025)\citenamefont
  {Nishimura}, \citenamefont {Funato}, \citenamefont {Matsuo},\ and\
  \citenamefont {Kato}}]{Nishimura2025}%
  \BibitemOpen
  \bibfield  {author} {\bibinfo {author} {\bibfnamefont {N.}~\bibnamefont
  {Nishimura}}, \bibinfo {author} {\bibfnamefont {T.}~\bibnamefont {Funato}},
  \bibinfo {author} {\bibfnamefont {M.}~\bibnamefont {Matsuo}},\ and\ \bibinfo
  {author} {\bibfnamefont {T.}~\bibnamefont {Kato}},\ }\bibfield  {title}
  {\bibinfo {title} {Theory of spin Seebeck effect activated by acoustic chiral
  phonons},\ }\href
  {https://doi.org/https://doi.org/10.1016/j.jmmm.2025.173386} {\bibfield
  {journal} {\bibinfo  {journal} {J. of Magn. Magn. Mater.}\ }\textbf {\bibinfo
  {volume} {630}},\ \bibinfo {pages} {173386} (\bibinfo {year}
  {2025})}\BibitemShut {NoStop}%
\bibitem [{\citenamefont {Yokoyama}(2026)}]{Yokoyama2026}%
  \BibitemOpen
  \bibfield  {author} {\bibinfo {author} {\bibfnamefont {T.}~\bibnamefont
  {Yokoyama}},\ }\bibfield  {title} {\bibinfo {title} {Supercurrent-induced
  phonon angular momentum},\ }\href@noop {} {\  (\bibinfo {year} {2026})},\
  \Eprint {https://arxiv.org/abs/2604.12701} {arXiv:2604.12701
  [cond-mat.mes-hall]} \BibitemShut {NoStop}%
\bibitem [{\citenamefont {Berry}(1984)}]{Berry1984}%
  \BibitemOpen
  \bibfield  {author} {\bibinfo {author} {\bibfnamefont {M.~V.}\ \bibnamefont
  {Berry}},\ }\bibfield  {title} {\bibinfo {title} {Quantal phase factors
  accompanying adiabatic changes},\ }\href
  {https://doi.org/https://doi.org/10.1098/rspa.1984.0023} {\bibfield
  {journal} {\bibinfo  {journal} {Proc. R. Soc. Lond. A}\ }\textbf {\bibinfo
  {volume} {392}},\ \bibinfo {pages} {45} (\bibinfo {year} {1984})}\BibitemShut
  {NoStop}%
\bibitem [{\citenamefont {Trifunovic}\ \emph {et~al.}(2019)\citenamefont
  {Trifunovic}, \citenamefont {Ono},\ and\ \citenamefont
  {Watanabe}}]{Trifunovic2019}%
  \BibitemOpen
  \bibfield  {author} {\bibinfo {author} {\bibfnamefont {L.}~\bibnamefont
  {Trifunovic}}, \bibinfo {author} {\bibfnamefont {S.}~\bibnamefont {Ono}},\
  and\ \bibinfo {author} {\bibfnamefont {H.}~\bibnamefont {Watanabe}},\
  }\bibfield  {title} {\bibinfo {title} {Geometric orbital magnetization in
  adiabatic processes},\ }\href {https://doi.org/10.1103/PhysRevB.100.054408}
  {\bibfield  {journal} {\bibinfo  {journal} {Phys. Rev. B}\ }\textbf {\bibinfo
  {volume} {100}},\ \bibinfo {pages} {054408} (\bibinfo {year}
  {2019})}\BibitemShut {NoStop}%
\bibitem [{\citenamefont {Yao}\ and\ \citenamefont {Murakami}(2022)}]{Yao2022}%
  \BibitemOpen
  \bibfield  {author} {\bibinfo {author} {\bibfnamefont {D.}~\bibnamefont
  {Yao}}\ and\ \bibinfo {author} {\bibfnamefont {S.}~\bibnamefont {Murakami}},\
  }\bibfield  {title} {\bibinfo {title} {Chiral-phonon-induced current in
  helical crystals},\ }\href {https://doi.org/10.1103/PhysRevB.105.184412}
  {\bibfield  {journal} {\bibinfo  {journal} {Phys. Rev. B}\ }\textbf {\bibinfo
  {volume} {105}},\ \bibinfo {pages} {184412} (\bibinfo {year}
  {2022})}\BibitemShut {NoStop}%
\bibitem [{\citenamefont {Hamada}\ and\ \citenamefont
  {Murakami}(2020)}]{Hamada2020}%
  \BibitemOpen
  \bibfield  {author} {\bibinfo {author} {\bibfnamefont {M.}~\bibnamefont
  {Hamada}}\ and\ \bibinfo {author} {\bibfnamefont {S.}~\bibnamefont
  {Murakami}},\ }\bibfield  {title} {\bibinfo {title} {Conversion between
  electron spin and microscopic atomic rotation},\ }\href
  {https://doi.org/10.1103/PhysRevResearch.2.023275} {\bibfield  {journal}
  {\bibinfo  {journal} {Phys. Rev. Res.}\ }\textbf {\bibinfo {volume} {2}},\
  \bibinfo {pages} {023275} (\bibinfo {year} {2020})}\BibitemShut {NoStop}%
\bibitem [{\citenamefont {Ren}\ \emph {et~al.}(2024)\citenamefont {Ren},
  \citenamefont {Bonini}, \citenamefont {Stengel}, \citenamefont {Dreyer},\
  and\ \citenamefont {Vanderbilt}}]{Ren2024}%
  \BibitemOpen
  \bibfield  {author} {\bibinfo {author} {\bibfnamefont {S.}~\bibnamefont
  {Ren}}, \bibinfo {author} {\bibfnamefont {J.}~\bibnamefont {Bonini}},
  \bibinfo {author} {\bibfnamefont {M.}~\bibnamefont {Stengel}}, \bibinfo
  {author} {\bibfnamefont {C.~E.}\ \bibnamefont {Dreyer}},\ and\ \bibinfo
  {author} {\bibfnamefont {D.}~\bibnamefont {Vanderbilt}},\ }\bibfield  {title}
  {\bibinfo {title} {Adiabatic dynamics of coupled spins and phonons in
  magnetic insulators},\ }\href {https://doi.org/10.1103/PhysRevX.14.011041}
  {\bibfield  {journal} {\bibinfo  {journal} {Phys. Rev. X}\ }\textbf {\bibinfo
  {volume} {14}},\ \bibinfo {pages} {011041} (\bibinfo {year}
  {2024})}\BibitemShut {NoStop}%
\bibitem [{\citenamefont {Yao}\ and\ \citenamefont {Murakami}(2024)}]{Yao2024}%
  \BibitemOpen
  \bibfield  {author} {\bibinfo {author} {\bibfnamefont {D.}~\bibnamefont
  {Yao}}\ and\ \bibinfo {author} {\bibfnamefont {S.}~\bibnamefont {Murakami}},\
  }\bibfield  {title} {\bibinfo {title} {Conversion of chiral phonons into
  magnons in ferromagnets and antiferromagnets},\ }\href
  {https://doi.org/10.7566/JPSJ.93.034708} {\bibfield  {journal} {\bibinfo
  {journal} {J. Phys. Soc. Jpn.}\ }\textbf {\bibinfo {volume} {93}},\ \bibinfo
  {pages} {034708} (\bibinfo {year} {2024})}\BibitemShut {NoStop}%
\bibitem [{\citenamefont {Yao}\ and\ \citenamefont {Murakami}(2025)}]{Yao2025}%
  \BibitemOpen
  \bibfield  {author} {\bibinfo {author} {\bibfnamefont {D.}~\bibnamefont
  {Yao}}\ and\ \bibinfo {author} {\bibfnamefont {S.}~\bibnamefont {Murakami}},\
  }\bibfield  {title} {\bibinfo {title} {Theory of spin magnetization driven by
  chiral phonons},\ }\href {https://doi.org/10.1103/PhysRevB.111.134414}
  {\bibfield  {journal} {\bibinfo  {journal} {Phys. Rev. B}\ }\textbf {\bibinfo
  {volume} {111}},\ \bibinfo {pages} {134414} (\bibinfo {year}
  {2025})}\BibitemShut {NoStop}%
\bibitem [{\citenamefont {Royo}\ and\ \citenamefont
  {Stengel}(2026)}]{Royo2025}%
  \BibitemOpen
  \bibfield  {author} {\bibinfo {author} {\bibfnamefont {M.}~\bibnamefont
  {Royo}}\ and\ \bibinfo {author} {\bibfnamefont {M.}~\bibnamefont {Stengel}},\
  }\bibfield  {title} {\bibinfo {title} {Dynamical response of noncollinear
  spin systems at constrained magnetic moments},\ }\href
  {https://doi.org/10.1103/c68v-8tkb} {\bibfield  {journal} {\bibinfo
  {journal} {Phys. Rev. X}\ }\textbf {\bibinfo {volume} {16}},\ \bibinfo
  {pages} {011049} (\bibinfo {year} {2026})}\BibitemShut {NoStop}%
\bibitem [{\citenamefont {Ren}\ \emph {et~al.}(2021)\citenamefont {Ren},
  \citenamefont {Xiao}, \citenamefont {Saparov},\ and\ \citenamefont
  {Niu}}]{Ren2021}%
  \BibitemOpen
  \bibfield  {author} {\bibinfo {author} {\bibfnamefont {Y.}~\bibnamefont
  {Ren}}, \bibinfo {author} {\bibfnamefont {C.}~\bibnamefont {Xiao}}, \bibinfo
  {author} {\bibfnamefont {D.}~\bibnamefont {Saparov}},\ and\ \bibinfo {author}
  {\bibfnamefont {Q.}~\bibnamefont {Niu}},\ }\bibfield  {title} {\bibinfo
  {title} {Phonon magnetic moment from electronic topological magnetization},\
  }\href {https://doi.org/10.1103/PhysRevLett.127.186403} {\bibfield  {journal}
  {\bibinfo  {journal} {Phys. Rev. Lett.}\ }\textbf {\bibinfo {volume} {127}},\
  \bibinfo {pages} {186403} (\bibinfo {year} {2021})}\BibitemShut {NoStop}%
\bibitem [{\citenamefont {Xiao}\ \emph {et~al.}(2021)\citenamefont {Xiao},
  \citenamefont {Ren},\ and\ \citenamefont {Xiong}}]{Xiao2021}%
  \BibitemOpen
  \bibfield  {author} {\bibinfo {author} {\bibfnamefont {C.}~\bibnamefont
  {Xiao}}, \bibinfo {author} {\bibfnamefont {Y.}~\bibnamefont {Ren}},\ and\
  \bibinfo {author} {\bibfnamefont {B.}~\bibnamefont {Xiong}},\ }\bibfield
  {title} {\bibinfo {title} {Adiabatically induced orbital magnetization},\
  }\href {https://doi.org/10.1103/PhysRevB.103.115432} {\bibfield  {journal}
  {\bibinfo  {journal} {Phys. Rev. B}\ }\textbf {\bibinfo {volume} {103}},\
  \bibinfo {pages} {115432} (\bibinfo {year} {2021})}\BibitemShut {NoStop}%
\bibitem [{\citenamefont {Yao}\ \emph {et~al.}(2025)\citenamefont {Yao},
  \citenamefont {Go}, \citenamefont {Mokrousov},\ and\ \citenamefont
  {Murakami}}]{YaoOAM2025}%
  \BibitemOpen
  \bibfield  {author} {\bibinfo {author} {\bibfnamefont {D.}~\bibnamefont
  {Yao}}, \bibinfo {author} {\bibfnamefont {D.}~\bibnamefont {Go}}, \bibinfo
  {author} {\bibfnamefont {Y.}~\bibnamefont {Mokrousov}},\ and\ \bibinfo
  {author} {\bibfnamefont {S.}~\bibnamefont {Murakami}},\ }\href@noop {}
  {\bibinfo {title} {Dynamical orbital angular momentum induced by circularly
  polarized phonons}} (\bibinfo {year} {2025}),\ \Eprint
  {https://arxiv.org/abs/arXiv:2511.09271} {arXiv:2511.09271} \BibitemShut
  {NoStop}%
\bibitem [{\citenamefont {Sato}\ \emph {et~al.}(2025)\citenamefont {Sato},
  \citenamefont {Kato},\ and\ \citenamefont {Manchon}}]{Sato2025}%
  \BibitemOpen
  \bibfield  {author} {\bibinfo {author} {\bibfnamefont {T.}~\bibnamefont
  {Sato}}, \bibinfo {author} {\bibfnamefont {T.}~\bibnamefont {Kato}},\ and\
  \bibinfo {author} {\bibfnamefont {A.}~\bibnamefont {Manchon}},\ }\href@noop
  {} {\bibinfo {title} {Orbital accumulation induced by chiral phonons}}
  (\bibinfo {year} {2025}),\ \Eprint {https://arxiv.org/abs/arXiv:2511.11272}
  {arXiv:2511.11272} \BibitemShut {NoStop}%
\bibitem [{\citenamefont {Pezo}\ \emph {et~al.}(2026)\citenamefont {Pezo},
  \citenamefont {Manchon}, \citenamefont {Nii}, \citenamefont {Ando},\ and\
  \citenamefont {Kato}}]{Pezp2026}%
  \BibitemOpen
  \bibfield  {author} {\bibinfo {author} {\bibfnamefont {A.}~\bibnamefont
  {Pezo}}, \bibinfo {author} {\bibfnamefont {A.}~\bibnamefont {Manchon}},
  \bibinfo {author} {\bibfnamefont {Y.}~\bibnamefont {Nii}}, \bibinfo {author}
  {\bibfnamefont {K.}~\bibnamefont {Ando}},\ and\ \bibinfo {author}
  {\bibfnamefont {T.}~\bibnamefont {Kato}},\ }\href@noop {} {\bibinfo {title}
  {First-principles prediction of chiral-phonon-induced orbital accumulation}}
  (\bibinfo {year} {2026}),\ \Eprint {https://arxiv.org/abs/arXiv:2605.03486}
  {arXiv:2605.03486} \BibitemShut {NoStop}%
\bibitem [{\citenamefont {Floquet}(1883)}]{Floquet1883}%
  \BibitemOpen
  \bibfield  {author} {\bibinfo {author} {\bibfnamefont {G.}~\bibnamefont
  {Floquet}},\ }\bibfield  {title} {\bibinfo {title} {On linear differential
  equations with periodic coefficients},\ }\href
  {https://doi.org/10.24033/asens.220} {\bibfield  {journal} {\bibinfo
  {journal} {Ann. Sci. \'Ec. Norm. Sup.}\ }\textbf {\bibinfo {volume} {12}},\
  \bibinfo {pages} {47} (\bibinfo {year} {1883})}\BibitemShut {NoStop}%
\bibitem [{\citenamefont {Shirley}(1965)}]{Shirley1965}%
  \BibitemOpen
  \bibfield  {author} {\bibinfo {author} {\bibfnamefont {J.~H.}\ \bibnamefont
  {Shirley}},\ }\bibfield  {title} {\bibinfo {title} {Solution of the
  Schr\"odinger equation with a Hamiltonian periodic in time},\ }\href
  {https://doi.org/10.1103/PhysRev.138.B979} {\bibfield  {journal} {\bibinfo
  {journal} {Phys. Rev.}\ }\textbf {\bibinfo {volume} {138}},\ \bibinfo {pages}
  {B979} (\bibinfo {year} {1965})}\BibitemShut {NoStop}%
\bibitem [{\citenamefont {Oka}\ and\ \citenamefont {Aoki}(2009)}]{Oka2009}%
  \BibitemOpen
  \bibfield  {author} {\bibinfo {author} {\bibfnamefont {T.}~\bibnamefont
  {Oka}}\ and\ \bibinfo {author} {\bibfnamefont {H.}~\bibnamefont {Aoki}},\
  }\bibfield  {title} {\bibinfo {title} {Photovoltaic Hall effect in
  graphene},\ }\href {https://doi.org/10.1103/PhysRevB.79.081406} {\bibfield
  {journal} {\bibinfo  {journal} {Phys. Rev. B}\ }\textbf {\bibinfo {volume}
  {79}},\ \bibinfo {pages} {081406(R)} (\bibinfo {year} {2009})}\BibitemShut
  {NoStop}%
\bibitem [{\citenamefont {Kitagawa}\ \emph {et~al.}(2010)\citenamefont
  {Kitagawa}, \citenamefont {Berg}, \citenamefont {Rudner},\ and\ \citenamefont
  {Demler}}]{Kitagawa2010}%
  \BibitemOpen
  \bibfield  {author} {\bibinfo {author} {\bibfnamefont {T.}~\bibnamefont
  {Kitagawa}}, \bibinfo {author} {\bibfnamefont {E.}~\bibnamefont {Berg}},
  \bibinfo {author} {\bibfnamefont {M.}~\bibnamefont {Rudner}},\ and\ \bibinfo
  {author} {\bibfnamefont {E.}~\bibnamefont {Demler}},\ }\bibfield  {title}
  {\bibinfo {title} {Topological characterization of periodically driven
  quantum systems},\ }\href {https://doi.org/10.1103/PhysRevB.82.235114}
  {\bibfield  {journal} {\bibinfo  {journal} {Phys. Rev. B}\ }\textbf {\bibinfo
  {volume} {82}},\ \bibinfo {pages} {235114} (\bibinfo {year}
  {2010})}\BibitemShut {NoStop}%
\bibitem [{\citenamefont {Kitagawa}\ \emph {et~al.}(2011)\citenamefont
  {Kitagawa}, \citenamefont {Oka}, \citenamefont {Brataas}, \citenamefont
  {Fu},\ and\ \citenamefont {Demler}}]{Kitagawa2011}%
  \BibitemOpen
  \bibfield  {author} {\bibinfo {author} {\bibfnamefont {T.}~\bibnamefont
  {Kitagawa}}, \bibinfo {author} {\bibfnamefont {T.}~\bibnamefont {Oka}},
  \bibinfo {author} {\bibfnamefont {A.}~\bibnamefont {Brataas}}, \bibinfo
  {author} {\bibfnamefont {L.}~\bibnamefont {Fu}},\ and\ \bibinfo {author}
  {\bibfnamefont {E.}~\bibnamefont {Demler}},\ }\bibfield  {title} {\bibinfo
  {title} {Transport properties of nonequilibrium systems under the application
  of light: Photoinduced quantum Hall insulators without Landau levels},\
  }\href {https://doi.org/10.1103/PhysRevB.84.235108} {\bibfield  {journal}
  {\bibinfo  {journal} {Phys. Rev. B}\ }\textbf {\bibinfo {volume} {84}},\
  \bibinfo {pages} {235108} (\bibinfo {year} {2011})}\BibitemShut {NoStop}%
\bibitem [{\citenamefont {Lindner}\ \emph {et~al.}(2011)\citenamefont
  {Lindner}, \citenamefont {Refael},\ and\ \citenamefont
  {Galitski}}]{Lindner2011}%
  \BibitemOpen
  \bibfield  {author} {\bibinfo {author} {\bibfnamefont {N.~H.}\ \bibnamefont
  {Lindner}}, \bibinfo {author} {\bibfnamefont {G.}~\bibnamefont {Refael}},\
  and\ \bibinfo {author} {\bibfnamefont {V.}~\bibnamefont {Galitski}},\
  }\bibfield  {title} {\bibinfo {title} {Floquet topological insulator in
  semiconductor quantum wells},\ }\href {https://doi.org/10.1038/nphys1926}
  {\bibfield  {journal} {\bibinfo  {journal} {Nature Physics}\ }\textbf
  {\bibinfo {volume} {7}},\ \bibinfo {pages} {490} (\bibinfo {year}
  {2011})}\BibitemShut {NoStop}%
\bibitem [{\citenamefont {Dag}\ and\ \citenamefont {Mitra}(2022)}]{Dag2022}%
  \BibitemOpen
  \bibfield  {author} {\bibinfo {author} {\bibfnamefont {C.~B.}\ \bibnamefont
  {Dag}}\ and\ \bibinfo {author} {\bibfnamefont {A.}~\bibnamefont {Mitra}},\
  }\bibfield  {title} {\bibinfo {title} {Floquet topological systems with flat
  bands: Edge modes, Berry curvature, and orbital magnetization},\ }\href
  {https://doi.org/10.1103/PhysRevB.105.245136} {\bibfield  {journal} {\bibinfo
   {journal} {Phys. Rev. B}\ }\textbf {\bibinfo {volume} {105}},\ \bibinfo
  {pages} {245136} (\bibinfo {year} {2022})}\BibitemShut {NoStop}%
\bibitem [{\citenamefont {Oka}\ and\ \citenamefont {Kitamura}(2019)}]{Oka2019}%
  \BibitemOpen
  \bibfield  {author} {\bibinfo {author} {\bibfnamefont {T.}~\bibnamefont
  {Oka}}\ and\ \bibinfo {author} {\bibfnamefont {S.}~\bibnamefont {Kitamura}},\
  }\bibfield  {title} {\bibinfo {title} {Floquet engineering of quantum
  materials},\ }\href
  {https://doi.org/https://doi.org/10.1146/annurev-conmatphys-031218-013423}
  {\bibfield  {journal} {\bibinfo  {journal} {Annu. Rev. Condens. Matter
  Phys.}\ }\textbf {\bibinfo {volume} {10}},\ \bibinfo {pages} {387} (\bibinfo
  {year} {2019})}\BibitemShut {NoStop}%
\bibitem [{\citenamefont {Rudner}\ and\ \citenamefont
  {Lindner}(2020)}]{Rudner2020}%
  \BibitemOpen
  \bibfield  {author} {\bibinfo {author} {\bibfnamefont {M.~S.}\ \bibnamefont
  {Rudner}}\ and\ \bibinfo {author} {\bibfnamefont {N.~H.}\ \bibnamefont
  {Lindner}},\ }\bibfield  {title} {\bibinfo {title} {Band structure
  engineering and non-equilibrium dynamics in Floquet topological insulators},\
  }\href {https://doi.org/10.1038/s42254-020-0170-z} {\bibfield  {journal}
  {\bibinfo  {journal} {Nature Reviews Physics}\ }\textbf {\bibinfo {volume}
  {2}},\ \bibinfo {pages} {229} (\bibinfo {year} {2020})}\BibitemShut {NoStop}%
\bibitem [{\citenamefont {Harper}\ \emph {et~al.}(2020)\citenamefont {Harper},
  \citenamefont {Roy}, \citenamefont {Rudner},\ and\ \citenamefont
  {Sondhi}}]{Harper2020}%
  \BibitemOpen
  \bibfield  {author} {\bibinfo {author} {\bibfnamefont {F.}~\bibnamefont
  {Harper}}, \bibinfo {author} {\bibfnamefont {R.}~\bibnamefont {Roy}},
  \bibinfo {author} {\bibfnamefont {M.~S.}\ \bibnamefont {Rudner}},\ and\
  \bibinfo {author} {\bibfnamefont {S.}~\bibnamefont {Sondhi}},\ }\bibfield
  {title} {\bibinfo {title} {Topology and broken symmetry in Floquet systems},\
  }\href
  {https://doi.org/https://doi.org/10.1146/annurev-conmatphys-031218-013721}
  {\bibfield  {journal} {\bibinfo  {journal} {Annu. Rev. Condens. Matter
  Phys.}\ }\textbf {\bibinfo {volume} {11}},\ \bibinfo {pages} {345} (\bibinfo
  {year} {2020})}\BibitemShut {NoStop}%
\bibitem [{\citenamefont {Wang}\ \emph {et~al.}(2026)\citenamefont {Wang},
  \citenamefont {Cai}, \citenamefont {Tang}, \citenamefont {Lu}, \citenamefont
  {Chen}, \citenamefont {Sheng}, \citenamefont {Feng}, \citenamefont {Zhong},
  \citenamefont {Zhang}, \citenamefont {Yu},\ and\ \citenamefont
  {Zhou}}]{Wang2026}%
  \BibitemOpen
  \bibfield  {author} {\bibinfo {author} {\bibfnamefont {F.}~\bibnamefont
  {Wang}}, \bibinfo {author} {\bibfnamefont {X.}~\bibnamefont {Cai}}, \bibinfo
  {author} {\bibfnamefont {X.}~\bibnamefont {Tang}}, \bibinfo {author}
  {\bibfnamefont {J.}~\bibnamefont {Lu}}, \bibinfo {author} {\bibfnamefont
  {W.}~\bibnamefont {Chen}}, \bibinfo {author} {\bibfnamefont {T.}~\bibnamefont
  {Sheng}}, \bibinfo {author} {\bibfnamefont {R.}~\bibnamefont {Feng}},
  \bibinfo {author} {\bibfnamefont {H.}~\bibnamefont {Zhong}}, \bibinfo
  {author} {\bibfnamefont {H.}~\bibnamefont {Zhang}}, \bibinfo {author}
  {\bibfnamefont {P.}~\bibnamefont {Yu}},\ and\ \bibinfo {author}
  {\bibfnamefont {S.}~\bibnamefont {Zhou}},\ }\bibfield  {title} {\bibinfo
  {title} {Observation of Floquet-induced gap in graphene},\ }\bibfield
  {journal} {\bibinfo  {journal} {Nature Materials}\ }\href
  {https://doi.org/10.1038/s41563-026-02549-y} {10.1038/s41563-026-02549-y}
  (\bibinfo {year} {2026})\BibitemShut {NoStop}%
\bibitem [{\citenamefont {Sentef}\ \emph {et~al.}(2015)\citenamefont {Sentef},
  \citenamefont {Claassen}, \citenamefont {Kemper}, \citenamefont {Moritz},
  \citenamefont {Oka}, \citenamefont {Freericks},\ and\ \citenamefont
  {Devereaux}}]{Sentef2015}%
  \BibitemOpen
  \bibfield  {author} {\bibinfo {author} {\bibfnamefont {M.~A.}\ \bibnamefont
  {Sentef}}, \bibinfo {author} {\bibfnamefont {M.}~\bibnamefont {Claassen}},
  \bibinfo {author} {\bibfnamefont {A.~F.}\ \bibnamefont {Kemper}}, \bibinfo
  {author} {\bibfnamefont {B.}~\bibnamefont {Moritz}}, \bibinfo {author}
  {\bibfnamefont {T.}~\bibnamefont {Oka}}, \bibinfo {author} {\bibfnamefont
  {J.~K.}\ \bibnamefont {Freericks}},\ and\ \bibinfo {author} {\bibfnamefont
  {T.~P.}\ \bibnamefont {Devereaux}},\ }\bibfield  {title} {\bibinfo {title}
  {Theory of Floquet band formation and local pseudospin textures in pump-probe
  photoemission of graphene},\ }\href {https://doi.org/10.1038/ncomms8047}
  {\bibfield  {journal} {\bibinfo  {journal} {Nature Communications}\ }\textbf
  {\bibinfo {volume} {6}},\ \bibinfo {pages} {7047} (\bibinfo {year}
  {2015})}\BibitemShut {NoStop}%
\bibitem [{\citenamefont {Kibis}(2022)}]{Kibis2022}%
  \BibitemOpen
  \bibfield  {author} {\bibinfo {author} {\bibfnamefont {O.~V.}\ \bibnamefont
  {Kibis}},\ }\bibfield  {title} {\bibinfo {title} {Floquet theory of spin
  dynamics under circularly polarized light pulses},\ }\href
  {https://doi.org/10.1103/PhysRevA.105.043106} {\bibfield  {journal} {\bibinfo
   {journal} {Phys. Rev. A}\ }\textbf {\bibinfo {volume} {105}},\ \bibinfo
  {pages} {043106} (\bibinfo {year} {2022})}\BibitemShut {NoStop}%
\bibitem [{\citenamefont {Neufeld}\ \emph {et~al.}(2023)\citenamefont
  {Neufeld}, \citenamefont {Tancogne-Dejean}, \citenamefont {De~Giovannini},
  \citenamefont {H{\"u}bener},\ and\ \citenamefont {Rubio}}]{Neufeld2023}%
  \BibitemOpen
  \bibfield  {author} {\bibinfo {author} {\bibfnamefont {O.}~\bibnamefont
  {Neufeld}}, \bibinfo {author} {\bibfnamefont {N.}~\bibnamefont
  {Tancogne-Dejean}}, \bibinfo {author} {\bibfnamefont {U.}~\bibnamefont
  {De~Giovannini}}, \bibinfo {author} {\bibfnamefont {H.}~\bibnamefont
  {H{\"u}bener}},\ and\ \bibinfo {author} {\bibfnamefont {A.}~\bibnamefont
  {Rubio}},\ }\bibfield  {title} {\bibinfo {title} {Attosecond magnetization
  dynamics in non-magnetic materials driven by intense femtosecond lasers},\
  }\href {https://doi.org/10.1038/s41524-023-00997-7} {\bibfield  {journal}
  {\bibinfo  {journal} {npj Computational Materials}\ }\textbf {\bibinfo
  {volume} {9}},\ \bibinfo {pages} {39} (\bibinfo {year} {2023})}\BibitemShut
  {NoStop}%
\bibitem [{\citenamefont {Tanaka}\ and\ \citenamefont
  {Sato}(2024)}]{Tanaka2024}%
  \BibitemOpen
  \bibfield  {author} {\bibinfo {author} {\bibfnamefont {M.}~\bibnamefont
  {Tanaka}}\ and\ \bibinfo {author} {\bibfnamefont {M.}~\bibnamefont {Sato}},\
  }\bibfield  {title} {\bibinfo {title} {Theory of the inverse Faraday effect
  in dissipative Rashba electron systems: Floquet engineering perspective},\
  }\href {https://doi.org/10.1103/PhysRevB.110.045204} {\bibfield  {journal}
  {\bibinfo  {journal} {Phys. Rev. B}\ }\textbf {\bibinfo {volume} {110}},\
  \bibinfo {pages} {045204} (\bibinfo {year} {2024})}\BibitemShut {NoStop}%
\bibitem [{\citenamefont {Murakami}\ \emph {et~al.}(2017)\citenamefont
  {Murakami}, \citenamefont {Tsuji}, \citenamefont {Eckstein},\ and\
  \citenamefont {Werner}}]{Murakami2017}%
  \BibitemOpen
  \bibfield  {author} {\bibinfo {author} {\bibfnamefont {Y.}~\bibnamefont
  {Murakami}}, \bibinfo {author} {\bibfnamefont {N.}~\bibnamefont {Tsuji}},
  \bibinfo {author} {\bibfnamefont {M.}~\bibnamefont {Eckstein}},\ and\
  \bibinfo {author} {\bibfnamefont {P.}~\bibnamefont {Werner}},\ }\bibfield
  {title} {\bibinfo {title} {Nonequilibrium steady states and transient
  dynamics of conventional superconductors under phonon driving},\ }\href
  {https://doi.org/10.1103/PhysRevB.96.045125} {\bibfield  {journal} {\bibinfo
  {journal} {Phys. Rev. B}\ }\textbf {\bibinfo {volume} {96}},\ \bibinfo
  {pages} {045125} (\bibinfo {year} {2017})}\BibitemShut {NoStop}%
\bibitem [{\citenamefont {Shin}\ \emph {et~al.}(2018)\citenamefont {Shin},
  \citenamefont {H{\"u}bener}, \citenamefont {De~Giovannini}, \citenamefont
  {Jin}, \citenamefont {Rubio},\ and\ \citenamefont {Park}}]{Shin2018}%
  \BibitemOpen
  \bibfield  {author} {\bibinfo {author} {\bibfnamefont {D.}~\bibnamefont
  {Shin}}, \bibinfo {author} {\bibfnamefont {H.}~\bibnamefont {H{\"u}bener}},
  \bibinfo {author} {\bibfnamefont {U.}~\bibnamefont {De~Giovannini}}, \bibinfo
  {author} {\bibfnamefont {H.}~\bibnamefont {Jin}}, \bibinfo {author}
  {\bibfnamefont {A.}~\bibnamefont {Rubio}},\ and\ \bibinfo {author}
  {\bibfnamefont {N.}~\bibnamefont {Park}},\ }\bibfield  {title} {\bibinfo
  {title} {Phonon-driven spin-Floquet magneto-valleytronics in MoS$_2$},\ }\href
  {https://doi.org/10.1038/s41467-018-02918-5} {\bibfield  {journal} {\bibinfo
  {journal} {Nature Communications}\ }\textbf {\bibinfo {volume} {9}},\
  \bibinfo {pages} {638} (\bibinfo {year} {2018})}\BibitemShut {NoStop}%
\bibitem [{\citenamefont {H{\"u}bener}\ \emph {et~al.}(2018)\citenamefont
  {H{\"u}bener}, \citenamefont {De~Giovannini},\ and\ \citenamefont
  {Rubio}}]{Hubener2018}%
  \BibitemOpen
  \bibfield  {author} {\bibinfo {author} {\bibfnamefont {H.}~\bibnamefont
  {H{\"u}bener}}, \bibinfo {author} {\bibfnamefont {U.}~\bibnamefont
  {De~Giovannini}},\ and\ \bibinfo {author} {\bibfnamefont {A.}~\bibnamefont
  {Rubio}},\ }\bibfield  {title} {\bibinfo {title} {Phonon driven Floquet
  matter},\ }\href {https://doi.org/10.1021/acs.nanolett.7b05391} {\bibfield
  {journal} {\bibinfo  {journal} {Nano Letters}\ }\textbf {\bibinfo {volume}
  {18}},\ \bibinfo {pages} {1535} (\bibinfo {year} {2018})}\BibitemShut
  {NoStop}%
\bibitem [{\citenamefont {Chaudhary}\ \emph {et~al.}(2020)\citenamefont
  {Chaudhary}, \citenamefont {Haim}, \citenamefont {Peng},\ and\ \citenamefont
  {Refael}}]{Chaudhary2020}%
  \BibitemOpen
  \bibfield  {author} {\bibinfo {author} {\bibfnamefont {S.}~\bibnamefont
  {Chaudhary}}, \bibinfo {author} {\bibfnamefont {A.}~\bibnamefont {Haim}},
  \bibinfo {author} {\bibfnamefont {Y.}~\bibnamefont {Peng}},\ and\ \bibinfo
  {author} {\bibfnamefont {G.}~\bibnamefont {Refael}},\ }\bibfield  {title}
  {\bibinfo {title} {Phonon-induced Floquet topological phases protected by
  space-time symmetries},\ }\href
  {https://doi.org/10.1103/PhysRevResearch.2.043431} {\bibfield  {journal}
  {\bibinfo  {journal} {Phys. Rev. Res.}\ }\textbf {\bibinfo {volume} {2}},\
  \bibinfo {pages} {043431} (\bibinfo {year} {2020})}\BibitemShut {NoStop}%
\bibitem [{\citenamefont {Klebl}\ \emph {et~al.}(2025)\citenamefont {Klebl},
  \citenamefont {Schobert}, \citenamefont {Eckstein}, \citenamefont
  {Sangiovanni}, \citenamefont {Balatsky},\ and\ \citenamefont
  {Wehling}}]{Klebl2025}%
  \BibitemOpen
  \bibfield  {author} {\bibinfo {author} {\bibfnamefont {L.}~\bibnamefont
  {Klebl}}, \bibinfo {author} {\bibfnamefont {A.}~\bibnamefont {Schobert}},
  \bibinfo {author} {\bibfnamefont {M.}~\bibnamefont {Eckstein}}, \bibinfo
  {author} {\bibfnamefont {G.}~\bibnamefont {Sangiovanni}}, \bibinfo {author}
  {\bibfnamefont {A.~V.}\ \bibnamefont {Balatsky}},\ and\ \bibinfo {author}
  {\bibfnamefont {T.~O.}\ \bibnamefont {Wehling}},\ }\bibfield  {title}
  {\bibinfo {title} {Ultrafast pseudomagnetic fields from electron-nuclear
  quantum geometry},\ }\href {https://doi.org/10.1103/PhysRevLett.134.016705}
  {\bibfield  {journal} {\bibinfo  {journal} {Phys. Rev. Lett.}\ }\textbf
  {\bibinfo {volume} {134}},\ \bibinfo {pages} {016705} (\bibinfo {year}
  {2025})}\BibitemShut {NoStop}%
\bibitem [{\citenamefont {Haldane}(1988)}]{Haldane1985}%
  \BibitemOpen
  \bibfield  {author} {\bibinfo {author} {\bibfnamefont {F.~D.~M.}\
  \bibnamefont {Haldane}},\ }\bibfield  {title} {\bibinfo {title} {Model for a
  quantum Hall effect without Landau levels: Condensed-matter realization of
  the "parity anomaly"},\ }\href {https://doi.org/10.1103/PhysRevLett.61.2015}
  {\bibfield  {journal} {\bibinfo  {journal} {Phys. Rev. Lett.}\ }\textbf
  {\bibinfo {volume} {61}},\ \bibinfo {pages} {2015} (\bibinfo {year}
  {1988})}\BibitemShut {NoStop}%
\bibitem [{\citenamefont {Nova}\ \emph {et~al.}(2017)\citenamefont {Nova},
  \citenamefont {Cartella}, \citenamefont {Cantaluppi}, \citenamefont
  {F{\"o}rst}, \citenamefont {Bossini}, \citenamefont {Mikhaylovskiy},
  \citenamefont {Kimel}, \citenamefont {Merlin},\ and\ \citenamefont
  {Cavalleri}}]{Nova2017}%
  \BibitemOpen
  \bibfield  {author} {\bibinfo {author} {\bibfnamefont {T.~F.}\ \bibnamefont
  {Nova}}, \bibinfo {author} {\bibfnamefont {A.}~\bibnamefont {Cartella}},
  \bibinfo {author} {\bibfnamefont {A.}~\bibnamefont {Cantaluppi}}, \bibinfo
  {author} {\bibfnamefont {M.}~\bibnamefont {F{\"o}rst}}, \bibinfo {author}
  {\bibfnamefont {D.}~\bibnamefont {Bossini}}, \bibinfo {author} {\bibfnamefont
  {R.~V.}\ \bibnamefont {Mikhaylovskiy}}, \bibinfo {author} {\bibfnamefont
  {A.~V.}\ \bibnamefont {Kimel}}, \bibinfo {author} {\bibfnamefont
  {R.}~\bibnamefont {Merlin}},\ and\ \bibinfo {author} {\bibfnamefont
  {A.}~\bibnamefont {Cavalleri}},\ }\bibfield  {title} {\bibinfo {title} {An
  effective magnetic field from optically driven phonons},\ }\href
  {https://doi.org/10.1038/nphys3925} {\bibfield  {journal} {\bibinfo
  {journal} {Nature Physics}\ }\textbf {\bibinfo {volume} {13}},\ \bibinfo
  {pages} {132} (\bibinfo {year} {2017})}\BibitemShut {NoStop}%
\bibitem [{\citenamefont {Goldman}\ and\ \citenamefont
  {Dalibard}(2014)}]{Goldman2014}%
  \BibitemOpen
  \bibfield  {author} {\bibinfo {author} {\bibfnamefont {N.}~\bibnamefont
  {Goldman}}\ and\ \bibinfo {author} {\bibfnamefont {J.}~\bibnamefont
  {Dalibard}},\ }\bibfield  {title} {\bibinfo {title} {Periodically driven
  quantum systems: Effective Hamiltonians and engineered gauge fields},\ }\href
  {https://doi.org/10.1103/PhysRevX.4.031027} {\bibfield  {journal} {\bibinfo
  {journal} {Phys. Rev. X}\ }\textbf {\bibinfo {volume} {4}},\ \bibinfo {pages}
  {031027} (\bibinfo {year} {2014})}\BibitemShut {NoStop}%
\bibitem [{\citenamefont {Bukov}\ \emph {et~al.}(2015)\citenamefont {Bukov},
  \citenamefont {D'Alessio},\ and\ \citenamefont {Polkovnikov}}]{Bukov2015}%
  \BibitemOpen
  \bibfield  {author} {\bibinfo {author} {\bibfnamefont {M.}~\bibnamefont
  {Bukov}}, \bibinfo {author} {\bibfnamefont {L.}~\bibnamefont {D'Alessio}},\
  and\ \bibinfo {author} {\bibfnamefont {A.}~\bibnamefont {Polkovnikov}},\
  }\bibfield  {title} {\bibinfo {title} {Universal high-frequency behavior of
  periodically driven systems: from dynamical stabilization to Floquet
  engineering},\ }\href {https://doi.org/10.1080/00018732.2015.1055918}
  {\bibfield  {journal} {\bibinfo  {journal} {Advances in Physics}\ }\textbf
  {\bibinfo {volume} {64}},\ \bibinfo {pages} {139} (\bibinfo {year}
  {2015})}\BibitemShut {NoStop}%
\bibitem [{\citenamefont {Eckardt}(2017)}]{Eckardt2017}%
  \BibitemOpen
  \bibfield  {author} {\bibinfo {author} {\bibfnamefont {A.}~\bibnamefont
  {Eckardt}},\ }\bibfield  {title} {\bibinfo {title} {Colloquium: Atomic
  quantum gases in periodically driven optical lattices},\ }\href
  {https://doi.org/10.1103/RevModPhys.89.011004} {\bibfield  {journal}
  {\bibinfo  {journal} {Rev. Mod. Phys.}\ }\textbf {\bibinfo {volume} {89}},\
  \bibinfo {pages} {011004} (\bibinfo {year} {2017})}\BibitemShut {NoStop}%
\bibitem [{SM()}]{SM}%
  \BibitemOpen
  \href@noop {} {}\bibinfo {howpublished}
  {\url{URL_will_be_inserted_by_publisher}},\ \bibinfo {note} {see Supplemental
  Materials for detailed discussions and calculations}\BibitemShut {NoStop}%
\bibitem [{\citenamefont {Yan}\ \emph {et~al.}(2007)\citenamefont {Yan},
  \citenamefont {Zhang}, \citenamefont {Kim},\ and\ \citenamefont
  {Pinczuk}}]{Yan2007}%
  \BibitemOpen
  \bibfield  {author} {\bibinfo {author} {\bibfnamefont {J.}~\bibnamefont
  {Yan}}, \bibinfo {author} {\bibfnamefont {Y.}~\bibnamefont {Zhang}}, \bibinfo
  {author} {\bibfnamefont {P.}~\bibnamefont {Kim}},\ and\ \bibinfo {author}
  {\bibfnamefont {A.}~\bibnamefont {Pinczuk}},\ }\bibfield  {title} {\bibinfo
  {title} {Electric field effect tuning of electron-phonon coupling in
  graphene},\ }\href {https://doi.org/10.1103/PhysRevLett.98.166802} {\bibfield
   {journal} {\bibinfo  {journal} {Phys. Rev. Lett.}\ }\textbf {\bibinfo
  {volume} {98}},\ \bibinfo {pages} {166802} (\bibinfo {year}
  {2007})}\BibitemShut {NoStop}%
\bibitem [{\citenamefont {Xiao}\ \emph {et~al.}(2005)\citenamefont {Xiao},
  \citenamefont {Shi},\ and\ \citenamefont {Niu}}]{Xiao2005}%
  \BibitemOpen
  \bibfield  {author} {\bibinfo {author} {\bibfnamefont {D.}~\bibnamefont
  {Xiao}}, \bibinfo {author} {\bibfnamefont {J.}~\bibnamefont {Shi}},\ and\
  \bibinfo {author} {\bibfnamefont {Q.}~\bibnamefont {Niu}},\ }\bibfield
  {title} {\bibinfo {title} {Berry phase correction to electron density of
  states in solids},\ }\href {https://doi.org/10.1103/PhysRevLett.95.137204}
  {\bibfield  {journal} {\bibinfo  {journal} {Phys. Rev. Lett.}\ }\textbf
  {\bibinfo {volume} {95}},\ \bibinfo {pages} {137204} (\bibinfo {year}
  {2005})}\BibitemShut {NoStop}%
\bibitem [{\citenamefont {Thonhauser}\ \emph {et~al.}(2005)\citenamefont
  {Thonhauser}, \citenamefont {Ceresoli}, \citenamefont {Vanderbilt},\ and\
  \citenamefont {Resta}}]{Thonhauser2005}%
  \BibitemOpen
  \bibfield  {author} {\bibinfo {author} {\bibfnamefont {T.}~\bibnamefont
  {Thonhauser}}, \bibinfo {author} {\bibfnamefont {D.}~\bibnamefont
  {Ceresoli}}, \bibinfo {author} {\bibfnamefont {D.}~\bibnamefont
  {Vanderbilt}},\ and\ \bibinfo {author} {\bibfnamefont {R.}~\bibnamefont
  {Resta}},\ }\bibfield  {title} {\bibinfo {title} {Orbital magnetization in
  periodic insulators},\ }\href {https://doi.org/10.1103/PhysRevLett.95.137205}
  {\bibfield  {journal} {\bibinfo  {journal} {Phys. Rev. Lett.}\ }\textbf
  {\bibinfo {volume} {95}},\ \bibinfo {pages} {137205} (\bibinfo {year}
  {2005})}\BibitemShut {NoStop}%
\end{thebibliography}

\end{document}

\title{Supplemental Material for ``Magnetism and Topology from Circularly Polarized Phonon Floquet Engineering"}

\author{Dapeng Yao}
%\email{dapeng.yao@riken.jp}
\affiliation{RIKEN Center for Emergent Matter Science (CEMS), 2-1 Hirosawa, Wako, Saitama 351-0198, Japan}

\author{Tiantian Zhang}
\affiliation{Institute of Theoretical Physics, Chinese Academy of Sciences, Beijing 100190, China}

\author{Takashi Oka}
\affiliation{The Institute for Solid State Physics, The University of Tokyo, Kashiwa, Chiba 277-8581, Japan}

\author{Takehito Yokoyama}
\affiliation{Department of Physics, Institute of Science Tokyo, 2-12-1 Ookayama, Meguro-ku, Tokyo 152-8551, Japan}

\maketitle

%\appendix
%\maketitle

%\author{Name}
%\affiliation{affiliation}
\renewcommand{\theequation}{S\arabic{equation}}
\setcounter{equation}{0}
\renewcommand{\tablename}{Table S}
\setcounter{table}{0}
\renewcommand{\thefigure}{S\arabic{figure}}
\setcounter{figure}{0}
\renewcommand{\thesection}{\Roman{section}}

\setcounter{section}{0}

\onecolumngrid

%\tableofcontents

%\clearpage

\textit{\textbf{Effective Hamiltonian in Floquet picture.---}}
Here, we show the detailed derivation of the effective Floquet Hamiltonian in the main text.
We begin with the real-space Hamiltonian of electrons with the Rashba
spin-orbital coupling (SOC), which reads
\begin{align}\label{H0}
\hat{H}_0=t_0\sum_{\braket{ij}}\hat{c}_i^{\dagger}\hat{c}_j+\sum_{i}\delta_i\hat{c}^{\dagger}_i\hat{c}_i+i\lambda_{\text{R}}\sum_{\braket{ij}}\hat{c}_i^{\dagger}\left(\bm s\times\frac{\bm d_{ij}}{|\bm d_{ij}|}\right)_z\hat{c}_j
\end{align}
where $\hat{c}_i=(\hat{c}_{i\uparrow},\hat{c}_{i\downarrow})^{\mathsf{T}}$ $[\hat{c}^{\dagger}_i=(\hat{c}^{\dagger}_{i\uparrow},\hat{c}^{\dagger}_{i\downarrow})]$ is the annihilation (creation) operator of electron at the $i$th site. When circularly polarized phonons are present, the dynamical Hamiltonian describing the electron-phonon coupling reads
\begin{align}
\hat{H}_{\text{ep}}(t)=&-\frac{t_0}{a_0^2}\sum_{\braket{ij}}[\bm u_{ij}(t)\cdot\bm d_{ij}]\hat{c}_i^{\dagger}\hat{c}_j+i\frac{\lambda_{\text{R}}}{a_0}\sum_{\braket{ij}}\hat{c}_i^{\dagger}[\bm s\times\bm u_{ij}(t)]_z\hat{c}_j\nonumber\\
&-i\frac{\lambda_{\text{R}}}{a_0^3}\sum_{\braket{ij}}[\bm u_{ij}(t)\cdot\bm d_{ij}]\hat{c}_i^{\dagger}(\bm s\times\bm d_{ij})_z\hat{c}_j,
\end{align}
where we take the counterclockwise mode as an example, and the time-dependent phonon relative displacement is given by $\bm u_{ij}(t)=u_{ij}(\cos\Omega t,\sin\Omega t)$ with $u_{ij}=u_j-u_i$ being the relative rotational amplitude.
Then the time-dependent Hamiltonian can be expended by 
\begin{align}
\hat{H}_{\text{ep}}(t)=\hat{H}_1e^{i\Omega t}+\hat{H}_{-1}e^{-i\Omega t}
\end{align}
with the Fourier mode
\begin{align}
\hat{H}_1=\sum_{\braket{ij}}J_{ij,0}^{(1)}\hat{c}_i^{\dagger}s_0\hat{c}_j+J_{ij,x}^{(1)}\hat{c}_i^{\dagger}s_x\hat{c}_j+J_{ij,y}^{(1)}\hat{c}_i^{\dagger}s_y\hat{c}_j,
\end{align}
where
\begin{align}
&J_{ij,0}^{(1)}=-\frac{t_0u_{ij}}{2a_0}e^{-i\theta_{ij}},\\
&J_{ij,x}^{(1)}=\frac{\lambda_{\text{R}}u_{ij}}{4a_0}(1+e^{-i2\theta_{ij}}),\\
&J_{ij,y}^{(1)}=-i\frac{\lambda_{\text{R}}u_{ij}}{4a_0}(1-e^{-i2\theta_{ij}}),
\end{align}
with the azimuthal angle $\theta_{ij}=\text{arctan}(d_{ij}^y/d_{ij}^x)$ of the vector $\bm d_{ij}$. The Hermiticity of the Hamiltonian requires $\hat{H}_{-1}=\hat{H}_1^{\dagger}$. We can easily check the $J_{ij}^{(-1)}=(J_{ji}^{(1)})^*$ by using $u_{ji}=-u_{ij}$, and $\theta_{ji}=\theta_{ij}+\pi$ which leads to $e^{i\theta_{ji}}=-e^{i\theta_{ij}}$ and $e^{i2\theta_{ji}}=e^{i2\theta_{ij}}$.
To obtain the effective Floquet Hamiltonian, we first calculate the following commutator:
\begin{align}\label{commu}
[\hat{H}_{-1},\hat{H}_{1}]&=\left[\sum_{\braket{ij}}\sum_{\alpha}J_{ij,\alpha}^{(-1)}\hat{c}_i^{\dagger}s_{\alpha}\hat{c}_j,\sum_{\braket{kl}}\sum_{\beta}J_{kl,\beta}^{(1)}\hat{c}_k^{\dagger}s_{\beta}\hat{c}_l\right]\nonumber\\
&=\sum_{\braket{ij}}\sum_{\braket{kl}}\sum_{\alpha\beta}J_{ij,\alpha}^{(-1)}J_{kl,\beta}^{(1)}\left[\hat{c}_i^{\dagger}s_{\alpha}\hat{c}_j,\hat{c}_k^{\dagger}s_{\beta}\hat{c}_l\right],
\end{align}
where
\begin{align}
\left[\hat{c}_i^{\dagger}s_{\alpha}\hat{c}_j,\hat{c}_k^{\dagger}s_{\beta}\hat{c}_l\right]=&\sum_{\sigma\sigma'}\sum_{\rho\rho'}(s_{\alpha})_{\sigma\sigma'}(s_{\beta})_{\rho\rho'}\left[\hat{c}^{\dagger}_{i\sigma}\hat{c}_{j\sigma'},\hat{c}^{\dagger}_{k\rho}\hat{c}_{l\rho'}\right]\nonumber\\
=&\sum_{\sigma\sigma'}\sum_{\rho\rho'}(s_{\alpha})_{\sigma\sigma'}(s_{\beta})_{\rho\rho'}\left(\delta_{jk}\delta_{\sigma'\rho}\hat{c}_{i\sigma}^{\dagger}\hat{c}_{l\rho'}-\delta_{il}\delta_{\sigma\rho'}\hat{c}_{k\rho}^{\dagger}\hat{c}_{j\sigma'}\right)\nonumber\\
=&\sum_{\sigma\rho'}\left[\sum_{\sigma'}(s_{\alpha})_{\sigma\sigma'}(s_{\beta})_{\sigma'\rho'}\right]\delta_{jk}\hat{c}_{i\sigma}^{\dagger}\hat{c}_{l\rho'}-\sum_{\rho\sigma'}\left[\sum_{\sigma}(s_{\beta})_{\rho\sigma}(s_{\alpha})_{\sigma\sigma'}\right]\delta_{il}\hat{c}_{k\rho}^{\dagger}\hat{c}_{j\sigma'}\nonumber\\
=&\delta_{jk}\sum_{\sigma\rho'}\hat{c}_{i\sigma}^{\dagger}(s_{\alpha}s_{\beta})_{\sigma\rho'}\hat{c}_{l\rho'}-\delta_{il}\sum_{\rho\sigma'}\hat{c}_{k\rho}^{\dagger}(s_{\beta}s_{\alpha})_{\rho\sigma'}\hat{c}_{j\sigma'}\nonumber\\
=&\delta_{jk}\hat{c}_{i}^{\dagger}(s_{\alpha}s_{\beta})\hat{c}_l-\delta_{il}\hat{c}_{k}^{\dagger}(s_{\beta}s_{\alpha})\hat{c}_j.
\end{align}
Then the commutator in Eq.~(\ref{commu}) is given by
\begin{align}
[\hat{H}_{-1},\hat{H}_{1}]&=\sum_{\braket{\braket{ij}}}\sum_{\alpha\beta}\hat{c}_i^{\dagger}\left\{J_{ik,\alpha}^{(-1)}J_{kj,\beta}^{(1)}(s_{\alpha}s_{\beta})-J_{ik,\beta}^{(1)}J^{(-1)}_{kj,\alpha}(s_{\beta}s_{\alpha})\right\}\hat{c}_j\nonumber\\
&=\sum_{\braket{\braket{ij}}}M_{ij,0}\hat{c}_i^{\dagger}s_0\hat{c}_j+M_{ij,x}\hat{c}_i^{\dagger}s_x\hat{c}_j+M_{ij,y}\hat{c}_i^{\dagger}s_y\hat{c}_j+M_{ij,z}\hat{c}_i^{\dagger}s_z\hat{c}_j,
\end{align}
where $k$ denotes the intermediate nearest-neighbor site connecting the next-nearest-neighbor sites $i$ and $j$ through the two-step hopping process $i\rightarrow k\rightarrow j$, and these coefficients reads 
\begin{align}
M_{ij,0}&=i\sqrt{3}\nu_{ij}\left[\left(\frac{t_0u_r}{2a_0}\right)^2-2\left(\frac{\lambda_{\text{R}}u_r}{4a_0}\right)^2\right],\\
M_{ij,x}&=-\frac{t_0\lambda_{\text{R}}}{4}\left(\frac{u_r}{a_0}\right)^2\cos\frac{\theta_{ki}+\theta_{kj}}{2},\\
M_{ij,y}&=\frac{t_0\lambda_{\text{R}}}{4}\left(\frac{u_r}{a_0}\right)^2\sin\frac{\theta_{ki}+\theta_{kj}}{2},\\
M_{ij,z}&=2\left(\frac{\lambda_{\text{R}}u_r}{4a_0}\right)^2.
\end{align}
Therefore, the effective Floquet Hamiltonian is obtained by using the van-Vleck type expansion with respect to $\Omega^{-1}$~\cite{Kitagawa2011,Goldman2014,Bukov2015,Eckardt2017}:
\begin{align}
\hat{H}_{\text{F}}=\hat{H}_0+\frac{1}{2\hbar\Omega}[\hat{H}_{-1},\hat{H}_{1}],
\end{align}
where the second term reads
\begin{align}\label{H_Omega}
\frac{1}{2\hbar\Omega}[\hat{H}_{-1},\hat{H}_{1}]=&~\frac{\tilde{A}}{2\hbar\Omega}\sum_{\braket{\braket{ij}}}i\nu_{ij}\hat{c}_i^{\dagger}s_0\hat{c}_j-\frac{\tilde{B}}{2\hbar\Omega}\sum_{\braket{\braket{ij}}}\cos\frac{\theta_{ki}+\theta_{kj}}{2}\hat{c}_i^{\dagger}s_x\hat{c}_j\nonumber\\
&+\frac{\tilde{B}}{2\hbar\Omega}\sum_{\braket{\braket{ij}}}\sin\frac{\theta_{ki}+\theta_{kj}}{2}\hat{c}_i^{\dagger}s_y\hat{c}_j+\frac{\tilde{C}}{2\hbar\Omega}\sum_{\braket{\braket{ij}}}\hat{c}_i^{\dagger}s_z\hat{c}_j
\end{align}
with $\nu_{ij}$ taking $\pm1$ for the next-nearest-neighbor hopping with a clockwise (counterclockwise) path from the $j$th site to the $i$th site. Equation.~(\ref{H_Omega}) corresponds to Eq.~(4) in the main text. These coefficients are given by
\begin{align}
\tilde A\equiv\frac{\sqrt{3}}{8}(2t_0^2-\lambda_{\text{R}}^2)\left(\frac{u_r}{a_0}\right)^2,~\tilde B\equiv\frac{t_0\lambda_{\text{R}}}{4}\left(\frac{u_r}{a_0}\right)^2,~\tilde C\equiv\frac{\lambda_{\text{R}}^2}{8}\left(\frac{u_r}{a_0}\right)^2.
\end{align}

\textit{\textbf{Threefold rotation symmetry.---}}
The effective Floquet Hamiltonian in real space holds the threefold $C_{3z}$ rotation symmetry with respect to the $z$ axis. After $C_{3z}$ rotation operation, the azimuthal angles changes as $\theta_{ki}\rightarrow\theta_{ki}+\frac{2\pi}{3}$, leading to
\begin{align}
-\cos\frac{\theta_{ki}+\theta_{kj}}{2}s_x+\sin\frac{\theta_{ki}+\theta_{kj}}{2}s_y\rightarrow&-\cos\left(\frac{\theta_{ki}+\theta_{kj}}{2}+\frac{2\pi}{3}\right)s_x+\sin\left(\frac{\theta_{ki}+\theta_{kj}}{2}+\frac{2\pi}{3}\right)s_y\nonumber\\
=&-\cos\frac{\theta_{ki}+\theta_{kj}}{2}\left(-\frac{1}{2}s_x-\frac{\sqrt{3}}{2}s_y\right)+\sin\frac{\theta_{ki}+\theta_{kj}}{2}\left(\frac{\sqrt{3}}{2}s_x-\frac{1}{2}s_y\right)\nonumber\\
=&-\cos\frac{\theta_{ki}+\theta_{kj}}{2}s'_x+\sin\frac{\theta_{ki}+\theta_{kj}}{2}s'_y,
\end{align}
where $(s'_x,s'_y)=(-\frac{1}{2}s_x-\frac{\sqrt{3}}{2}s_y,\frac{\sqrt{3}}{2}s_x-\frac{1}{2}s_y)$ represents the spin rotation, and the other terms which do not depend on azimuthal angles are invariant. Therefore, the total Hamiltonian is invariant under the $C_{3z}$ rotation operation.

\textit{\textbf{Topological phase transition with Rashba SOC.---}}
In the absence of the Rashba SOC, the spin-up and spin-down electron bands remain doubly degenerate. When the system undergoes the topological phase transition with increasing the phonon rotational amplitude, the phonon dynamics yields a Haldane-type mass term, driving the system into a topological phase with Chern number $C=-2$ due to the spin degeneracy. When the Rashba SOC is present, the spin degeneracy is lifted, and the spin splitting further leads to a gap closing and reopening near the $K$ point by varying the Rashba SOC. 

\begin{figure}
\begin{center}
\includegraphics[width=15cm]{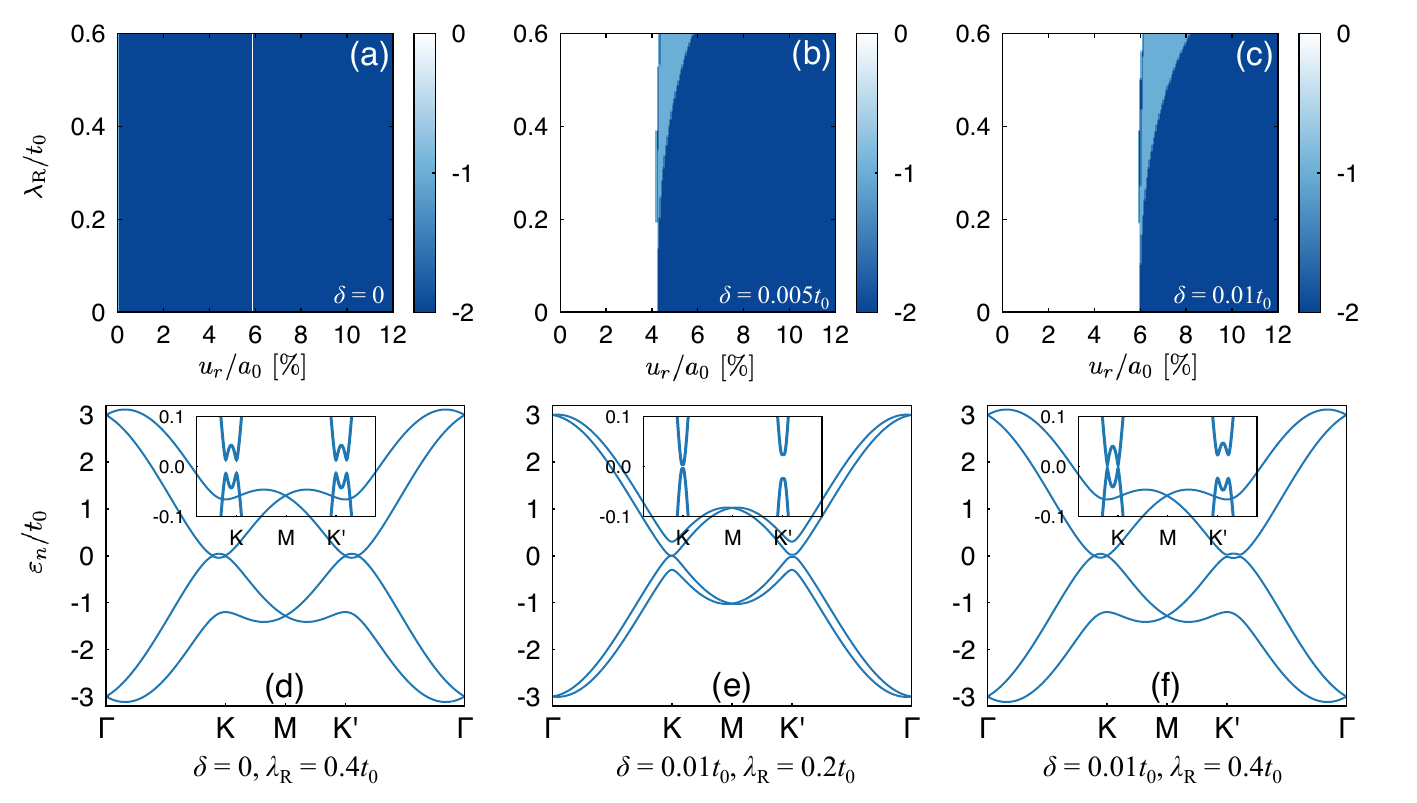}
\caption{Phase diagram of the Chern number as a function of the Rashba SOC $\lambda_{\text{R}}/t_0$ and the phonon rotational amplitude $u_r/a_0$ for different on-site potentials (a) $\delta=0$, (b) $\delta=0.005t_0$, and (c) $\delta=0.01t_0$. The band structures with spin splitting are shown for (d) $\delta=0,~\lambda_{\text{R}}=0.4t_0$, (e) $\delta=0.01t_0,~\lambda_{\text{R}}=0.2t_0$, and (e) $\delta=0.01t_0,~\lambda_{\text{R}}=0.4t_0$.
The phonon energy in (a-f) and phonon rotation amplitude in (d-f) are given by $\hbar\Omega=0.2t_0$ and $u_r/a_0=7\%$, respectively.}
\label{sm_fig1}
\end{center}
\end{figure}

Here we show the phase diagram for different on-site potentials as $\delta=0$, $\delta=0.005t_0$, and $\delta=0.01t_0$ in Figs.~\ref{sm_fig1}(a), \ref{sm_fig1}(b), and \ref{sm_fig1}(c), respectively.
Since the band structure without phonons is gapless when the on-site potential is zero $(\delta=0)$, the gap opens from the Haldane-type mass term as long as the circularly polarized phonon is present. We show the band structure without on-site potential for the Rashba SOC $\lambda_{\text{R}}=0.04t_0$ in Fig.~\ref{sm_fig1}(d). 
In the case of $\delta=0$, the spin splitting does not further lead to a gap closing, thereby the Chern number in the topological phase remains $C=-2$.
When the on-site potential is present, the band gaps near the $K$ and $K'$ points are different as shown in Figs.~\ref{sm_fig1}(e) and~\ref{sm_fig1}(f). As discussed in the main text, in the spinless case, the band gaps at the $K$ and $K'$ points is given by $\Delta_{\mp}=\delta\mp\frac{3\sqrt{3}\tilde{A}}{4\hbar\Omega}$. By increasing the Rashba SOC, the electron bands are spin-split, and the spin-splitting bands touch near the $K$ point. For example, by fixing the on-site potential $\delta=0.01t_0$, corresponding to the phase diagram in Fig.~\ref{sm_fig1}(c), when we change the Rashba SOC from $\lambda_{\text{R}}=0.2t_0$ to $\lambda_{\text{R}}=0.4t_0$, the band gap closes near the $K$ point accompanied by the change of Chern number from $C=-2$ to $C=-1$ as shown in Figs.~\ref{sm_fig1}(e) and ~\ref{sm_fig1}(f).

\textit{\textbf{Spin magnetization.---}}
The expectation value of electron spin is generally given by
\begin{align}
\braket{\bm S}=-i\int\frac{d\omega}{2\pi}\text{Tr}\left[\hat{\bm S}G^{<}(\omega)\right],
\end{align}
where $\hat{\bm S}$ denotes the spin-vector operator, and $G^{<}(\omega)=f(\omega)[G^{\text{A}}(\omega)-G^{\text{R}}(\omega)]$ represents the lesser Green's function with the Fermi-Dirac distribution function $f(\omega)$ and the retarded (advanced) Green's function $G^{\text{R}(\text{A})}(\omega)$.
Accordingly, the spectrum function is defined as
\begin{align}
A(\omega)\equiv i[G^{\text{R}}(\omega)-G^{\text{A}}(\omega)],
\end{align}
yielding $-iG^{<}(\omega)=f(\omega)A(\omega)$. Therefore, we can rewrite the expectation value of the electron spin in terms of the spectrum function as
\begin{align}
\braket{\bm S}=\int\frac{d\omega}{2\pi}f(\omega)\text{Tr}\left[\hat{\bm S}A(\omega)\right].
\end{align}
In the Bloch-band representation, the retarded (advanced) Green's function is given by
\begin{align}
G^{\text{R}(\text{A})}(\bm k,\omega)=\sum_n\frac{\ket{u_{n\bm k}}\bra{u_{n\bm k}}}{\omega-E_{n\bm k}\pm i0^+},
\end{align}
and therefore the spectrum function becomes
\begin{align}
A(\bm k,\omega)=2\pi\sum_n\ket{u_{n\bm k}}\bra{u_{n\bm k}}\delta(\omega-E_{n\bm k}).
\end{align}
Finally, the expectation value of $\hat{\bm S}$ in equilibrium is given by
\begin{align}
\braket{\bm S}=\int\frac{d^2k}{(2\pi)^2}\sum_n\braket{\hat{\bm S}}_{n\bm k}f(E_{n\bm k}),
\end{align}
where $\braket{\hat{\bm S}}_{n\bm k}\equiv\bra{u_{n\bm k}}\hat{\bm S}\ket{u_{n\bm k}}$ denotes the spin texture of the $n$th band in momentum space.